\documentclass{ametsocV6.1}

\title{Effects of Varying Land Coverage, Rotation Period, and Water Vapor on Equatorial Climates that Bridge the Gap between Earth-like and Titan-like}

\authors{M. M. McKinney,\aff{a}\correspondingauthor{M. M. McKinney, mmckinney@atmos.ucla.edu} 
J. Mitchell,\aff{a,b} 
S. I. Thomson,\aff{c}
}

\affiliation{\aff{a}{Atmospheric \& Oceanic Sciences, University of California, Los Angeles}\\
\aff{b}{Earth, Planetary, and Space Sciences, University of California, Los Angeles}\\
\aff{c}{College of Engineering, Mathematics and Physical Sciences, University of Exeter, Exeter EX4 4QE, UK}
}

\abstract{Saturn's largest moon, Titan, has an Earth-like volatile cycle, but with methane playing the role of water and surface liquid reservoirs geographically isolated at high latitudes. We recreate Titan's characteristic dry hydroclimate at the equator of an Earth-like climate model without seasons and with water as the condensable by varying a small set of planetary parameters. We use three observationally motivated criteria for Titan-like conditions at the equator: 1) the peak in surface specific humidity is not at the equator, despite it having the warmest annual-mean temperatures; 2) the vertical profile of specific humidity in the equatorial column is nearly constant through the lower troposphere; and 3) the relative humidity near the surface at the equator is significantly lower than saturation (lower than 60\%). We find that simply reducing the available water at the equator does not fully reproduce Titan-like conditions. We additionally vary the rotation period and volatility of water to mimic Titan's slower rotation and more abundant methane vapor. Longer rotation periods coupled with a dry equatorial surface meet fewer of the Titan-like criteria than equivalent experiments with shorter rotation periods. Experiments with higher volatility of water meet more criteria than those with lower volatility, with some of those with the highest volatility meeting all three, demonstrating that an Earth-like planet can display Titan-like climatology by changing only a few physical parameters.} 

\begin{document}

\maketitle

%
%
%


%
%

%
\section{Introduction}

The planet Earth is unique in the Solar System for having a surface dominated by a deep reservoir of a condensable liquid. A condensable liquid in the context of planetary science is one that can evaporate from the surface into the atmosphere, subsequently condense and/or solidify, and then fall back to the surface as precipitation. Earth's condensable liquid, water, is abundant enough and paired with significant enough topography such that over 70\% of its surface is an ocean multiple kilometers deep. Earth is not alone in having deposits of a condensable liquid on its surface; Saturn's moon Titan possesses a thick atmosphere and methane lakes that can evaporate to form clouds \citep{turtle2018titan} and rain \citep{turtle2011rapid}. Titan's surface is not dominated by methane, however. Its lakes of liquid methane are concentrated at its two poles \citep{lunine2009rivers} and in a broader polar ``wetlands'' extending to the midlatitudes \citep{lora2015titan, mitchell2016climate} while the rest of its surface is a vast desert centered on the equator. This presents an interesting contrast to Earth, which is often described as an ``aquaplanet'', meaning its surface and climate are dominated by deep, global oceans. Titan can be considered its natural counterpart, the ``terraplanet'', where the surface and climate are dominated by dry land as a single global continent with geographically isolated reservoirs of surface liquid.

Titan differs from Earth in other important ways. Its tropics are uniformly higher elevation than its midlatitudes and poles, as opposed to topography on Earth which has significant local variation and minimal dependence on latitude. Its rotation period is long, ~16 Earth-days, weakening the effects of the Coriolis Force. On Earth the Coriolis force causes moving air parcels and ocean currents to curve in their path. This can be seen in many common weather patterns, including hurricanes and baroclinic storms. The Coriolis Force is zero at the equator and sub-dominant in an area confined to ~10-20$^{\circ}$N/S latitude. This sub-dominant area can be referred to as the ``tropics'' with regards to atmospheric dynamics because the weak Coriolis Force -- and correspondingly small temperature gradient -- prevents the formation of most large-scale weather systems common to the rest of the planet \citep{Sobel2001}. This region is much broader on Titan due to its slower rotation (and to some extent its smaller size), so we can describe Titan as an ``all-tropics'' planet \citep{lora2015titan}. This means that its Intertropical Convergence Zone (ITCZ), the location where air rises in the summer hemisphere tropics, moves deep into the summer hemisphere (e.g. \citealp{mitchell2016climate}, and citations therein; \citealp{Faulk2017}; \citealp{Guendelman2018}; \citealp{Singh2019}; \citealp{Hill2021, Hill2022}). Its condensable, methane, is more volatile under Titan's surface conditions than water is under Earth's. The volatility of a liquid is a measure of how easily it evaporates at a given temperature and pressure; a more volatile liquid will evaporate more readily than a less volatile liquid under the same conditions. In Titan's atmosphere the result is higher specific humidities (i.e the mass of condensable vapor to the total mass of air) of methane in the troposphere than can be achieved on Earth with water.

Since Titan is an all-tropics planet, it may be assumed that polar air would be advected into the tropics on its way to the ITCZ. This would allow for saturated parcels to reach the tropics and set the local specific humidity to levels similar to those at the colder poles, which would be the location of last saturation for these air parcels. Simulations of Titan's global climate \citep{lora2015titan} suggest this does not occur, as the specific humidity in the tropics appears to be \textit{lower} than polar values \citep[Figure 6 of][]{adamkovics2016meridional, lora2017near}. Methane cannot condense out of the air moving horizontally away from the poles as the temperatures increase and the relative humidity (RH) of the parcels decreases. This necessitates an alternative hypothesis to explain the lower specific humidity in Titan's tropics relative to its poles. \citet{griffith2013titan} suggested that the tropical humidity is set by falling virga from above the lower troposphere. Virga is precipitation that evaporates before reaching the ground, effectively acting as a transfer of specific humidity from one level of the atmosphere to a lower one. There is an additional hypothesis that, when combined with downward vertical flow, virga creates a near-constant specific humidity profile in the lower troposphere, which was observed by the Cassini-Huygens mission \citep[Figure 2 therein]{niemann2005abundances}. At the same time, temperatures in the lower troposphere follow the dry adiabat, meaning relative humidity (RH) falls quickly as you approach the surface \citep[Figure 1 of][]{Tokano2006}. If this hypothesis is true it can be used as a proxy for a Titan-like climate. \citet{mitchell2016climate} alternatively suggest that slantwise, meridional-vertical motion from baroclinic, mid-latitude storms could source the tropics with low-specific-humidity air. The two hypotheses may be distinguishable in simulations, for instance by estimating the region of last saturation for equatorial parcels \citep{o2006stochastic}.

Recent work by \citet{fan2021reducing} showed that high global evaporative resistance can yield near-constant specific humidity in the lower troposphere even in an Earth-like Global Climate Model (GCM). They pointed to the presence of negative surface evaporation as the driver of this extremely dry climate state. Negative evaporation is generally only possible with a surface temperature inversion, but can be achieved much more easily when a high evaporative resistance is applied. One motivating question for this work is whether negative evaporation is necessary to produce the near-constant vertical profiles in specific humidity. 

A complementary approach for understanding the controlling factors of Titan-like and Earth-like hydroclimate states is to first define the qualities that distinguish them and study how an idealized climate system behaves as it transitions from one to the other. A number of parameters differ between the Earth and Titan climate systems. Our hypothesis is that four such parameters control the hydroclimate state: evaporative resistance, equatorial land fraction, rotation period, and the volatility of the condensable. In this paper we develop a suite of numerical experiments using an Earth-based climate model that span the parameter range bounded by Earth and Titan, and study the suite of experiments to evaluate under what conditions the Earth-like climate becomes Titan-like. In order to perform this analysis, we will first need to define what counts as "Titan-like". There are several key features of Titan's hydroclimate that have been observed and can be identified in our model runs. We choose three of these to use as our Titan-like criteria:
\begin{enumerate}
\item The highest annual-mean near-surface specific humidity is off the equator, which we define as poleward of 5$^{\circ}$N/S latitude (referred to herein as the ``OffEq'' criterion), based on observed near-surface specific humidity shown in Figure 6 of \citet{adamkovics2016meridional},
\item The vertical specific humidity at the equator is constant or nearly so (the ``ConVQ'' criterion), as shown in Figure 2 of \citet{niemann2005abundances}, and
\item The equatorial climate is significantly drier than an Earth-like aquaplanet. The Huygens probe measured RH values at the surface of around 50-60\% \citep[Figure 1]{niemann2005abundances, Tokano2006}, so experiments with equatorial near-surface RH at or below 60\% will meet the ``LowRH'' criterion.\end{enumerate}
We introduce the climate model used for the numerical experiments in Section 2 and outline the structure of our numerical experiments and our analysis methods in Section 3. In Section 4, we present and analyze the numerical experiments. We offer some discussion and further insights, plus next steps in Section 5. We summarize the main findings and conclude in Section 6.

\section{Model Description}
Isca is an open-source climate modeling framework, which allows the creation of atmospheric models over a wide-range of complexities \citep{vallis2018isca}. It is based on the Flexible Modeling System (FMS) developed at GFDL (www.gfdl.noaa.gov/fms/), but has been modified significantly to include multiple options for various parameterizations, as well as options to simulate various planets and moons.

\begin{figure}[h]
\hspace{0.3\textwidth}\includegraphics[width=0.4\textwidth]{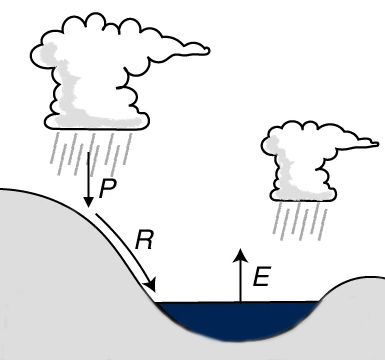}
\caption{Subsection taken from the diagram of the TAM hydrology module from \citet{faulk2020titan}. We use only the precipitation (P), runoff (R), and surface evaporation (E) from this scheme.}
\label{hydrofig}
\end{figure}

In the present work we use an idealized configuration of Isca, with mostly Earth-like parameters, but without any Earth-like continents, topography, or seasons. We mainly change the properties of the surface, which in the default Isca setup (referred to herein as ``Isca-default") is a mixed-layer ocean with a heat capacity equivalent to 40m of water. For our experiments we have incorporated a novel surface hydrology scheme developed by \citet{faulk2020titan} for the Titan Atmosphere Model \citep[TAM;][]{lora2015gcm}, referred to herein as ``Isca-hydro". This hydrology scheme treats the entire surface of the planet as ``land'', which can then have water on its surface. This water can run off into neighboring grid cells that are either at a lower elevation or lower relative water level, using the scheme shown in Figure \ref{hydrofig}. The water is not treated as a separate surface from the solid surface, and instead the model uses the depth of water in a grid cell to calculate the local heat capacity. As such there is only a single ``surface" for each grid cell, and all grid cells use the same calculations for surface parameters regardless of water content. The hydrology scheme from \citet{faulk2020titan} includes options for infiltration and subsurface flow, but we do not use any subsurface processes in our experiments.

The experiments are run using a suite of physics modules. These include a full Betts-Miller convection scheme \citep{betts1986new, betts1986newtwo}, the Socrates radiation scheme developed by the UK Met Office \citep{manners2017socrates, edwards1996studies}, and a simple large-scale condensation scheme. The settings for the Betts-Miller scheme are equivalent to those used in TAM from \citet{lora2015gcm} and \citet{faulk2020titan}, with the RH relaxing to 0.8 over a period of 7200 seconds. There is precipitation produced by the convection and condensation schemes, but there is no cloud module. Re-evaporation is included in the condensation scheme. For these experiments we use the same configuration of Socrates as in their ``Global Atmosphere 7'' configuration, with 6 bands in the short-wave and 9 in the long-wave \citep{manners2017socrates}, as described for Isca in \citet{thomson2019hierarchical}. We use a spectral dynamical core at T21 horizontal resolution (having 64 longitudinal gridpoints and 32 in latitude), and use 40 vertical sigma levels extending up to a top full-pressure level of 0.6hPa. The T21 resolution was chosen to minimize the time needed to run our experiments. We additionally ran a few sample experiments at T42 horizontal resolution to compare, and found similar mean states and convergence times.

The planet is based on Earth, with the same radius, solar constant, orbital period, 1 bar sea-level pressure in an N$_2$ atmosphere, and water as the condensable, but with 0$^{\circ}$ obliquity and eccentricity. The albedo is set to a constant value of 0.3 for all experiment setups and without distinguishing between land and water, which is to approximate Earth's mean albedo. The atmosphere is treated as Earth's pre-industrial atmosphere with CO\textsubscript{2}, a specified O\textsubscript{3} layer in the stratosphere, and radiative feedback from water vapor that enters the atmosphere through evaporation. We specify the O\textsubscript{3} field to be constant in time, and hemispherically symmetric, using the same setup as in \citet{jucker2017untangling}.

\section{Methods}
To make the baseline Earth-like planet of the model more Titan-like, we vary four parameters:
\begin{enumerate}
\item We start by mirroring the approach used in \citet{fan2021reducing}. For this we use an aquaplanet setup with an evaporative resistance, $r_E=100\times(1-\beta)$, where $\beta$ is a constant inserted into the equation for surface evaporation,
\begin{linenomath*}
\begin{equation}
E=\rho C_K L_v V_s (\beta q^*_s-q_1)
\end{equation}
\end{linenomath*}
where $\rho$ is the surface air density, $C_K$ is a transfer coefficient, $L_v$ is the latent heat of vaporization, $V_s$ is the surface wind speed, $q_s^*(T_s)$ is the saturation specific humidity at the surface temperature $T_s$, and $q_1$ is the lowest-model-layer specific humidity. We use the variable $r_E$, represented as a percentage, in this mansucript for convenience. This means that $r_E=100\%$ indicates a complete shutdown of evaporation and $r_E=0\%$ indicates the absence of evaporative resistance.

Evaporative resistance acts to reduce the surface evaporation without needing to create a dry surface in the model. All aquaplanet simulations have a global surface reservoir 40m deep, 1-day rotation, and water as the condensable. The hydrology scheme calculates evaporation using Equation 1 and specified values of $r_E$.

As previously mentioned, it is relatively easy to achieve negative evaporation when using a large value of $r_E$ (equivalent to a small, but nonzero, value of $\beta$) as the constant is only applied to one term of the difference. Due to the difficulty in connecting this to a physical phenomenon, we will attempt to produce a Titan-like state by varying other parameters that do not directly affect the evaporation equation.
\item We create a strip of land centered on the equator with a variable width in latitude, $\Delta\varphi$. This is accomplished using the Isca-hydro scheme which allows for custom global topography. By raising the solid surface only at a specified location and initializing an ocean elsewhere whose depth is just below the height of the raised area, we can create continents of varying widths. We use a Gaussian curve with a fixed peak elevation, 
\begin{linenomath*}
\begin{equation}
z(\varphi)=\left\{
    \begin{array}{ll}
    A\cos(f\varphi) & \quad \lvert\varphi\rvert\leq\Delta\varphi\\
    a\exp\left(\frac{-\varphi^2}{2c^2}\right)+1 & \quad \lvert\varphi\rvert>\Delta\varphi
    \end{array}
\right.
\end{equation}
\end{linenomath*}
where $z(\varphi)$ is the elevation of the solid surface as a function of latitude ($\varphi$), $A$ is the maximum elevation in meters (set to 70), and $\Delta\varphi$ is the latitude of the land strip's edge in both hemispheres. The coefficient $f$ is defined as a separate function of $\varphi$,
\begin{linenomath*}
\begin{equation}
f(\varphi)=\frac{\arcsin\left(\frac{60}{A}\right)-90}{\varphi}
\end{equation}
\end{linenomath*}
while the coefficients $a$ and $c$ are defined as functions of $\Delta\varphi$,
\begin{linenomath*}
\begin{equation}
a(\Delta\varphi)=\frac{60}{\exp\left(\frac{-\Delta\varphi^2}{2c^2}\right)}
\end{equation}
\end{linenomath*}
\begin{linenomath*}
\begin{equation}
c(\Delta\varphi)=\sqrt{\frac{-2\Delta\varphi-1}{2\ln(0.9)}}
\end{equation}
\end{linenomath*}
We use the following values for $\Delta\varphi$: 5$^{\circ}$, 15$^{\circ}$, 25$^{\circ}$, 35$^{\circ}$, 45$^{\circ}$, 55$^{\circ}$, 65$^{\circ}$, and 75$^{\circ}$. All $\Delta\varphi$ experiments use $r_E=0\%$, 1-day rotation, and water as the condensable. Grid cells outside of the land strip are initialized with 40m of water.

\item For experiments with $\Delta\varphi$ of 25$^{\circ}$, 35$^{\circ}$, and 45$^{\circ}$, we also vary the rotation period ($T_r$) to the following values in Earth-days: 1, 2, 4, 8, and 16. All $T_r$ experiments use $r_E=0\%$ and water as the condensable. Grid cells outside of the land strip are initialized with 40m of water. The upper value is chosen based on previous work by \citet{Faulk2017} who found that even at $T_r=8$ the ITCZ of an Earth-like planet could reach the poles.
\item For the same $\Delta\varphi$ in 3 we additionally vary $\xi$, defined as a constant factor applied to the Clausius-Clapeyron equation \citep[as in][]{Frierson2006}, such that
\begin{linenomath*}
\begin{equation}
e_s(T)=\xi e_0\exp\left[\frac{L_v}{R_v}\left(\frac{1}{T}-\frac{1}{T_0}\right)\right]
\end{equation}
\end{linenomath*}
where $e_s$ is the saturation vapor pressure as a function of temperature, $e_0$ is a known reference vapor pressure at temperature $T_0$, $L_v$ is the latent heat of vaporization, $R_v$ is the specific gas constant of water vapor, and $\xi$ takes the values 1, 1.25, 1.5, 1.75, 2, 2.25, and 2.5. The upper value is chosen so that the model does not enter a runaway greenhouse, since our model does include water vapor feedback. Due to this feedback, a doubling of $\xi$ results in at least one order-of-magnitude increase in column water vapor. Titan has two orders of magnitude more methane vapor in its atmosphere than Earth has water, much larger than we can achieve with these values of $\xi$. As we will show in Section 4, this range is sufficient to transition the Earth-like model to a Titan-like hydroclimate state. All $\xi$ experiments use $r_E=0\%$ and $T_r=1$. Grid cells outside of the land strip are initialized with 40m of water.
\end{enumerate}

To facilitate comparison between the equatorial hydroclimates of aquaplanet and terraplanet simulations, we diagnose the components of the zonal-mean atmospheric moisture budget due to large-scale condensation, convection, diffusion, and convergence of the moisture flux ($-\nabla \cdot F_Q$). The latter is further divided into mean ($-\nabla \cdot \overline{F_Q}$) and eddy components ($-\nabla \cdot F_Q'$), where
\begin{linenomath*}
\begin{equation}
-\nabla \cdot F_Q=-\nabla \cdot \overline{F_Q}-\nabla \cdot F_Q'
\end{equation}
\end{linenomath*}
\noindent We compare regions of moisture flux divergence and convergence to the effective surface of last saturation for equatorial, surface-level air, which we define as the dew point temperature of a parcel at the equatorial surface, $T_{d,eq}$. Overplotting the $T_{d,eq}$ isotherm on moisture flux divergence/ convergence allows us to inspect for equatorial surface moisture source regions.

\begin{figure}[h]
\includegraphics[width=1\textwidth]{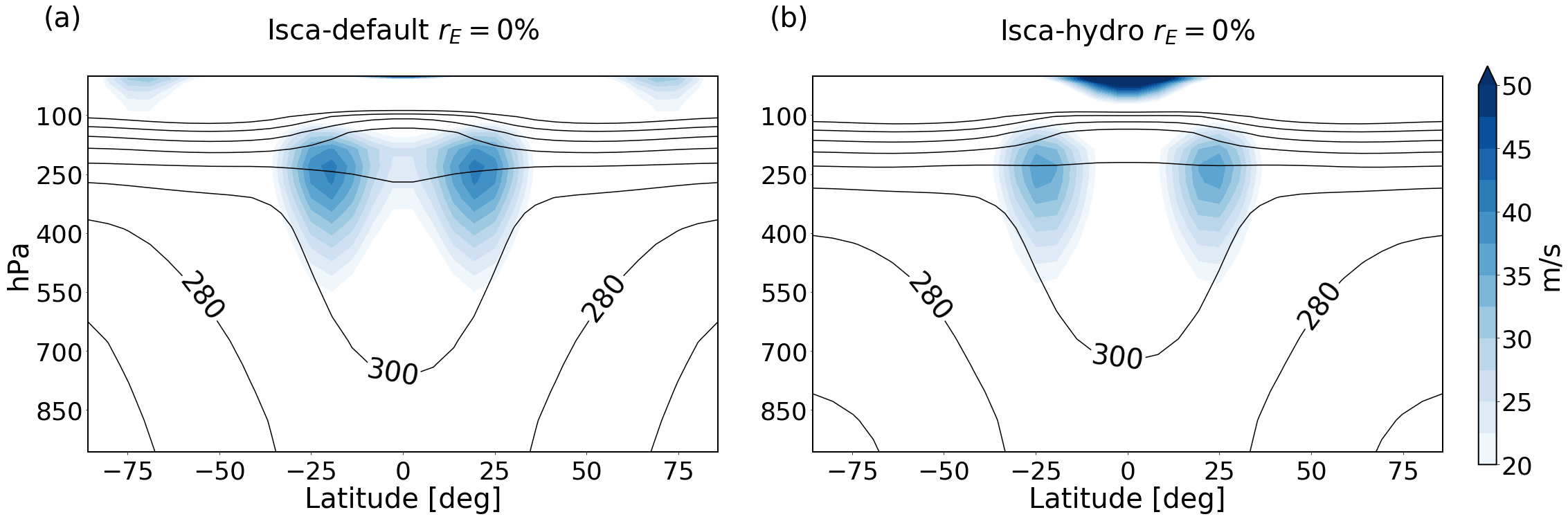}
\caption{Zonal- and time-mean zonal wind overlaid with the potential temperature field for (a) the Isca-default aquaplanet experiment with $r_E=0$ and (b) the equivalent Isca-hydro experiment. Filled contours show the zonal wind values in (m/s), while the black contour lines show the potential temperature in (K). Temperature contours are spaced 20K apart.}
\label{aqua_benchmarks}
\end{figure}

Simulations are performed for 25 years, and all diagnostics are averaged over the last 10 years. We observed that all experiments reached a steady state within 15 years, but do not demonstrate this herein for length considerations. Figure \ref{aqua_benchmarks} shows the zonal wind and potential temperature fields for two aquaplanet simulations with $r_E=0\%$ using the Isca-default (a) and Isca-hydro (b) surface schemes. This figure shows similar jet locations and peak winds for the two hydrology schemes with minor differences. The potential temperature values in our sample cases are somewhat cooler than those from \citet[subsequently HS]{Held1994}, but the shape of the distribution is similar. The jet locations are closer to the equator, around $25^\circ$ latitude, compared to approximately $40^\circ$ in HS. While the peak magnitudes of the jets are similar both between the two cases in Figure \ref{aqua_benchmarks} and the HS cases, we do find some superrotation at the equator. In the Isca-default case it is found in the upper troposphere, while in the Isca-hydro case it is in the stratosphere. Based on the comparison in Figure \ref{aqua_benchmarks}, we are confident that the Isca-hydro scheme is functional and effective for our analysis. Using model diagnostics of the hydrology, we then apply our three criteria for Titan-like climate states to assess whether a given parameter set has transitioned to a Titan-like hydroclimate.

\section{Results}
The results of our numerical experiments provide new insight into the transition from an Earth-like to Titan-like hydroclimate regime. We group the experiments by single parameter variations, i.e. keeping all others fixed.

\subsection{Variable $r_E$ Experiments}

\begin{figure}[h]

\includegraphics[width=1\textwidth]{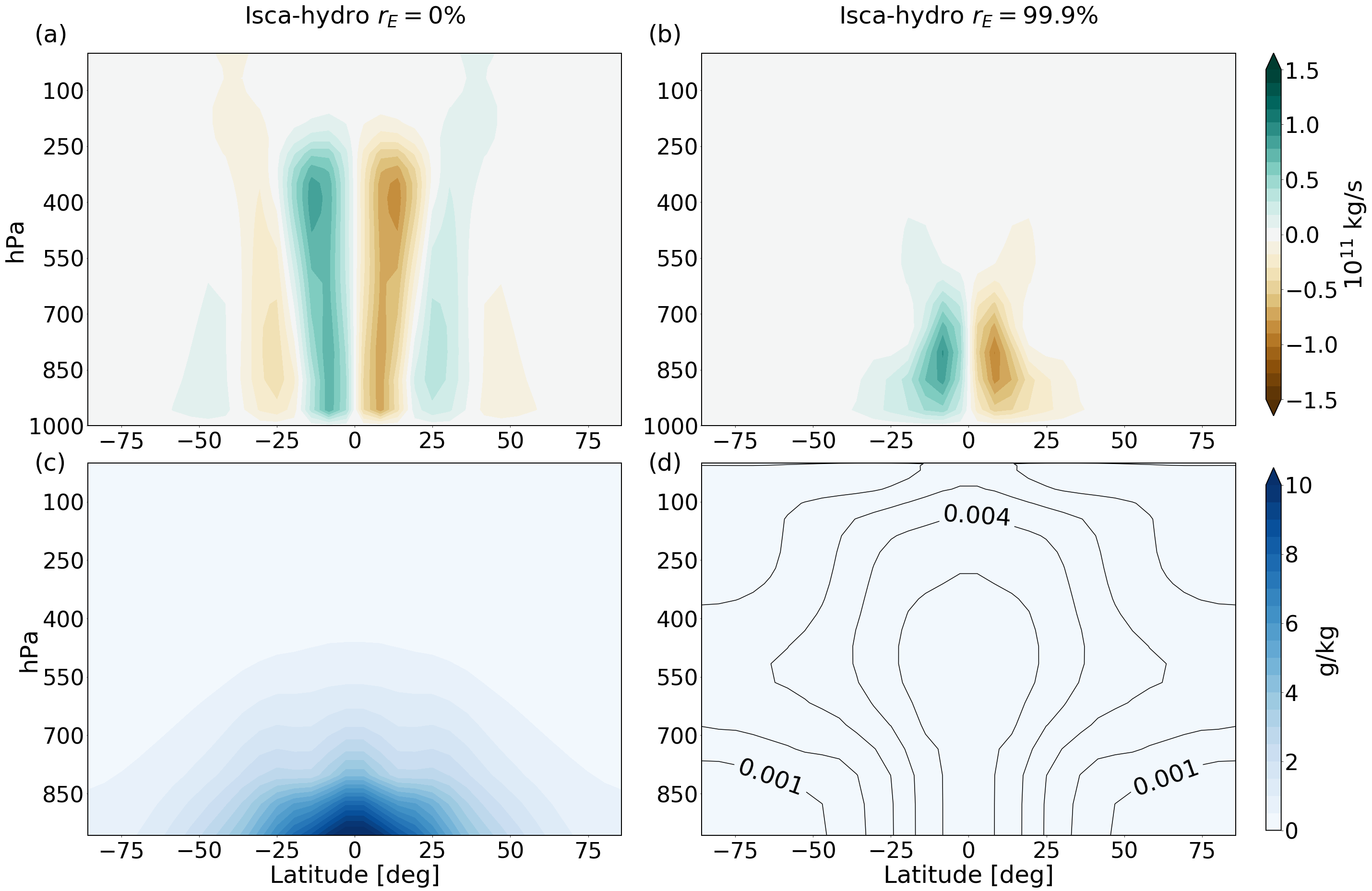}
\caption{Top row: Zonal- and time-mean stream function for the Isca-hydro aquaplanet experiments with (a) $r_E=0\%$ and (b) $r_E=99.9\%$. Bottom row: Equivalent to the top row but for the specific humidity fields of each experiment. In order to show the specific humidity distribution in (d), we include black contour lines showing values in g/kg and spaced 0.001 g/kg apart. We find a significant contraction in the vertical extent of the Hadley Cell (HC) with higher $r_E$, in accordance with an almost fully-dry atmosphere.}
\label{re_exp_basic_comp}
\end{figure}

We first run a set of aquaplanet experiments using the Isca-default hydrology scheme. These experiments are used to establish a baseline Earth-like state and the effect of a uniform evaporative resistance. In Figure \ref{re_exp_basic_comp} we take an initial look at the global circulation and specific humidity of the two end-member experiments for this experiment set, $r_E=0\%$ and $r_E=99.9\%$. The $r_E=0\%$ case has an Earth-like circulation (\ref{re_exp_basic_comp}(a)) and specific humidity field (\ref{re_exp_basic_comp}(c)), while the other case demonstrates a nearly-dry atmosphere (\ref{re_exp_basic_comp}(d)). Its Hadley Cell (HC) has contracted in the vertical dimension, extending only up to 600hPa (\ref{re_exp_basic_comp}(b)), indicating weak convection likely caused by the lack of moisture and subsequent latent heating.

\begin{figure}[h]

\includegraphics[width=1\textwidth]{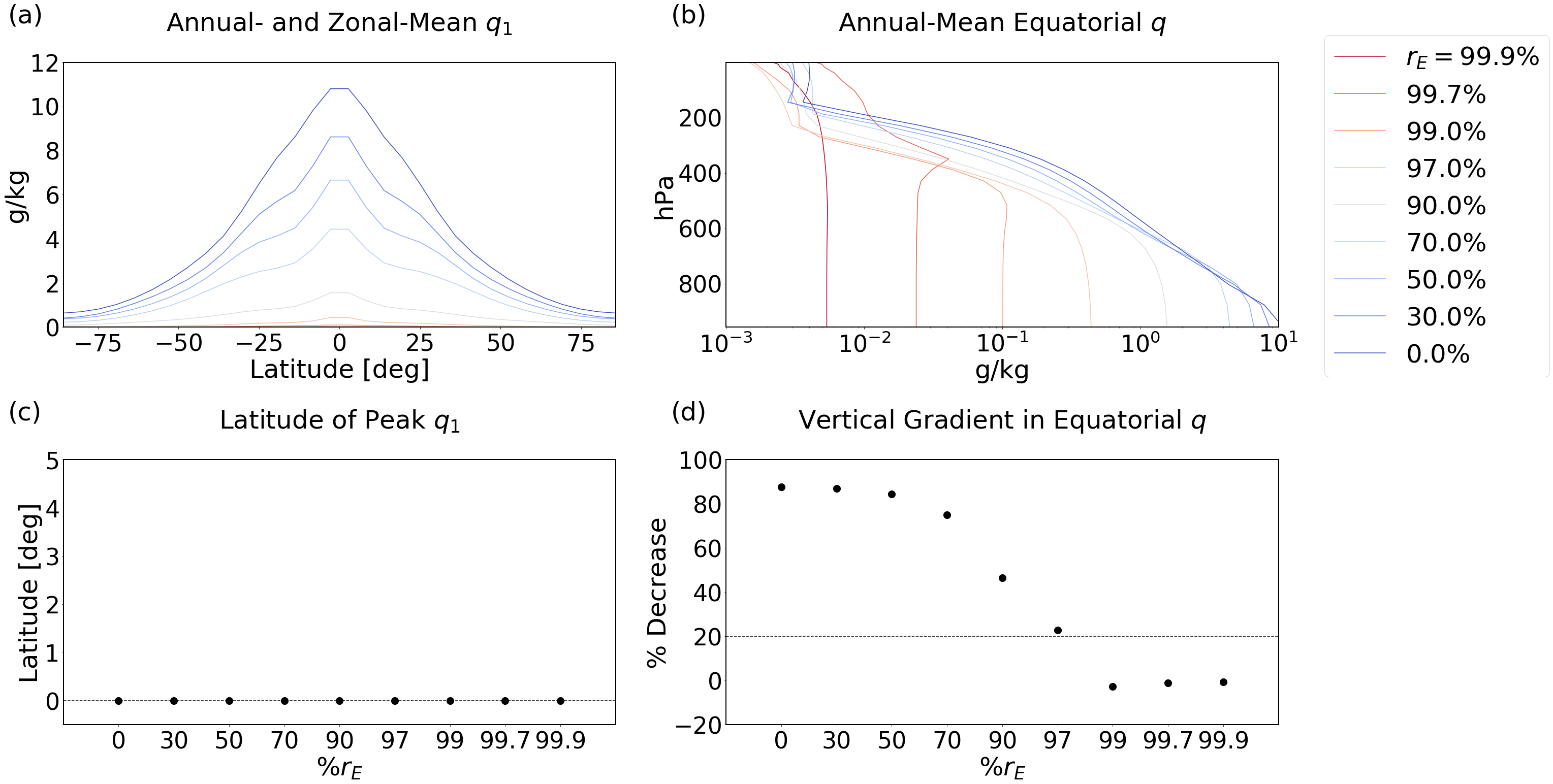}
\caption{Specific humidity analysis for the $r_E$ experiments. Figure (a) shows the zonal- and time-mean near-surface specific humidity, while (b) shows the time-mean vertical profile of specific humidity at the equator. Figure (c) shows the latitude of peak specific humidity from (a) as a function of $r_E$. The dotted line denotes $0^{\circ}$ latitude. Since all experiments have their peaks at $0^{\circ}$, none meet the OffEq criterion. Figure (d) shows the vertical gradient in specific humidity from (b) between the surface and 600hPa pressure level, calculated as a percent decrease from the surface value. The dotted line shows the 20\% threshold, and any experiments with gradients at or below that threshold meet the ConVQ criterion.}
\label{re_exp_sh}
\end{figure}

To assess the three Titan-like criteria, we look at the near-surface specific humidity. Figure \ref{re_exp_sh}(a) shows zonal- and time-mean $q_1$ for each $r_E$ experiment using the Isca-hydro scheme, while \ref{re_exp_sh}(c) shows the latitude of peak $q_1$ as a function of $r_E$ for the same experiments. All cases in the Isca-hydro setup have their peak humidity at the equator, failing to meet the OffEq criterion. Drier cases do have smaller gradients between the equator and poles, but since the $r_E$ parameter is applied uniformly everywhere on the surface the equator never becomes drier than the midlatitudes.

Next we assess the vertical profile of specific humidity for the equatorial column to see if any experiments meet the ConVQ criterion. Figure \ref{re_exp_sh}(b) shows the vertical profile of time-mean specific humidity at the equator for the $r_E$ experiments, while \ref{re_exp_sh}(d) shows the percent decrease in specific humidity moving from the surface to the 600hPa pressure level. The 20\% threshold to meet the ConVQ criterion is indicated by a dotted line. The percent decrease in specific humidity only begins to decrease once $r_E>50\%$, and only goes below the 20\% threshold in experiments with $r_E\geq99\%$. We conclude that the ConVQ criterion is only met for high values of $r_E$, which would indicate a very dry surface.

\begin{figure}[h]
\hspace{0.2\textwidth}\includegraphics[width=0.5\textwidth]{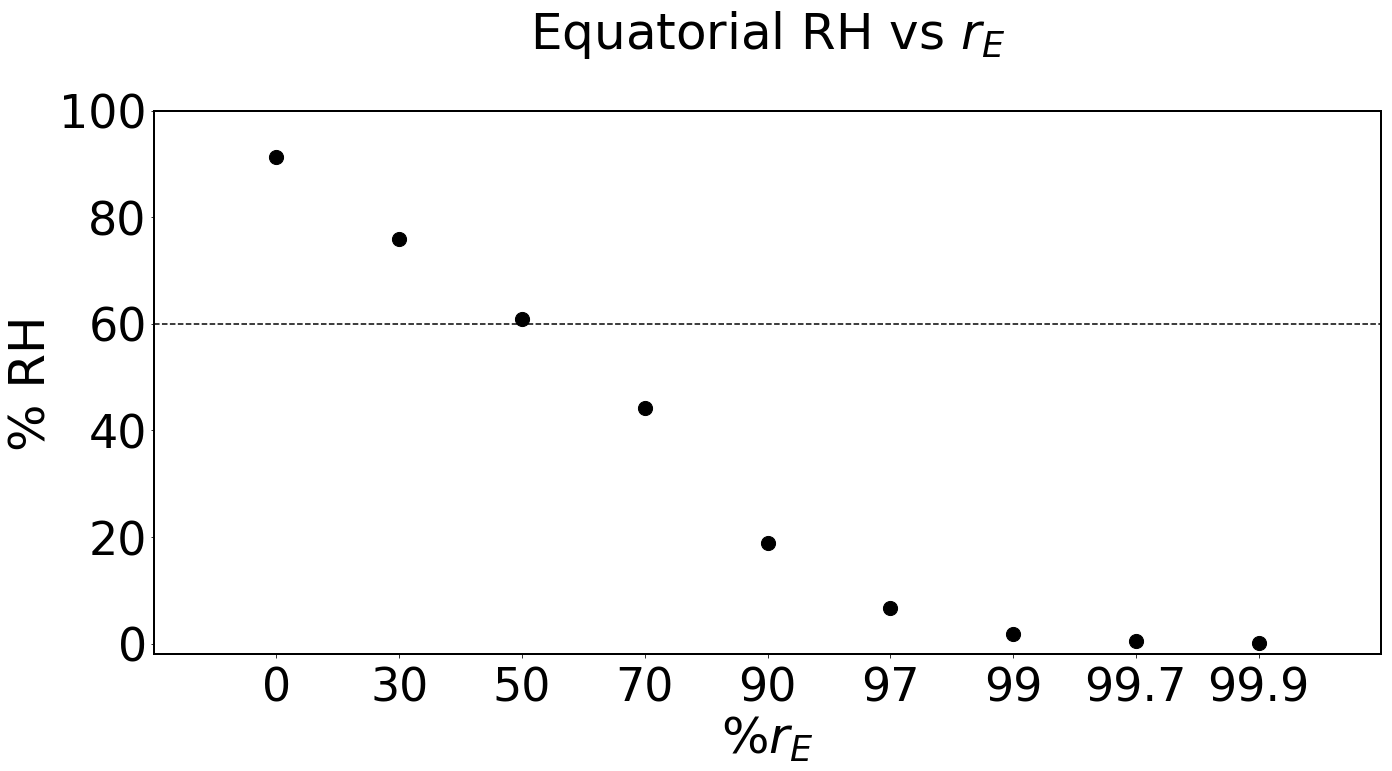}
\caption{Time-mean equatorial RH for the $r_E$ experiments as a function of $r_E$. The dotted line denotes 60\% RH, meaning experiments below it meet the LowRH criterion.}
\label{re_exp_rh}
\end{figure}

For the final criterion, LowRH, we show the equatorial time-mean RH for all $r_E$ experiments in Figure \ref{re_exp_rh}. RH drops quickly with increasing $r_E$, reaching values below 60\% for $r_E\geq70\%$. We conclude that experiments with $r_E\geq99\%$ meet both the ConVQ and LowRH criteria.

The results for the $r_E$ experiments are summarized in Table \ref{re_table}. We find that no experiments meet the OffEq criterion, but experiments with higher $r_E$ are more likely to meet the ConVQ and LowRH criteria. We infer that the $r_E$ experiments cannot meet the OffEq criterion because we apply a uniform $r_E$ value everywhere, whereas Titan is not uniformly dry over its surface. As such, we require a surface that includes a mix of dry and wet regions to develop more realism toward the Titan-like regime.

\begin{table}[h]
\begin{center}
\begin{tabular}{|c|c|c|c|}
\hline
$ $ & \multicolumn{3}{|c|}{Criteria}\\
\cline{2-4}
$ $ & OffEq & ConVQ & LowRH \\
\hline
$r_E=0\%$ & $ $ & $ $ & $ $\\
\hline
$30\%$ & $ $ & $ $ & $ $\\
\hline
$50\%$ & $ $ & $ $ & $ $\\
\hline
$70\%$ & $ $ & $ $ & X\\
\hline
$90\%$ & $ $ & $ $ & X\\
\hline
$97\%$ & $ $ & $ $ & X\\
\hline
$99\%$ & $ $ & X & X\\
\hline
$99.7\%$ & $ $ & X & X\\
\hline
$99.9\%$ & $ $ & X & X\\
\botline
\end{tabular}
\caption{Criteria matched by each $r_E$ experiment.}
\label{re_table}
\end{center}
\end{table}

\subsection{Variable $\Delta\varphi$ Experiments}

\begin{figure}[h]

\includegraphics[width=1\textwidth]{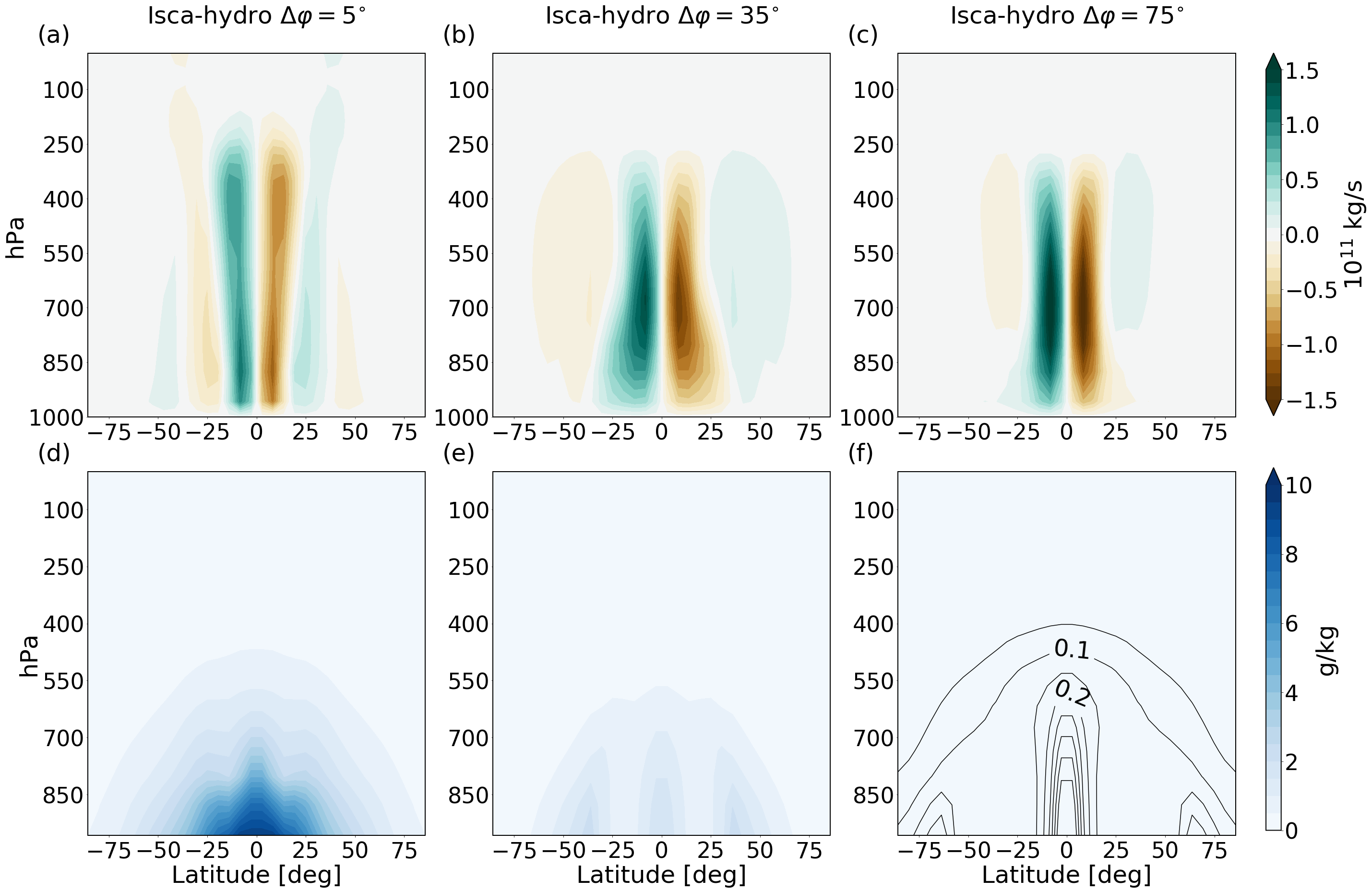}
\caption{Top row: Zonal- and time-mean stream function for the Isca-hydro experiments with (a) $\Delta\varphi=5^{\circ}$, (b) $\Delta\varphi=35^{\circ}$, and (c) $\Delta\varphi=75^{\circ}$. Bottom row: Equivalent to the top row but for the specific humidity fields of each experiment. As in Figure \ref{re_exp_basic_comp}(d), we add black contour lines in (f) to show the specific humidity distribution. These contours show values in g/kg and are spaced 0.05 g/kg apart. Similar to the effect of high $r_E$, we find the vertical extent of the HC decreases with larger $\Delta\varphi$ as the atmosphere becomes dry.}
\label{lat_exp_basic_comp}
\end{figure}

As detailed in Section 2, we add a hemispherically symmetric strip of land to the Isca-hydro scheme by introducing global topography with a moderately elevated equatorial region. The width of said strip is adjusted by adjusting the topography, and runoff is self-consistently simulated while surface infiltration is prohibited. We first analyze the effect of the land strip's width on the stream function and specific humidity fields in Figure \ref{lat_exp_basic_comp}. We find a similar vertical contraction and strengthening of the HC with higher $\Delta\varphi$ (\ref{lat_exp_basic_comp}(a-c)) as in Figure \ref{re_exp_basic_comp} with higher $r_E$. In both cases the driest end-member is almost completely dry relative to the first, meaning the presence of a large land strip has a comparable effect to high evaporative resistance, as might be expected. 

\begin{figure}[h]

\includegraphics[width=1\textwidth]{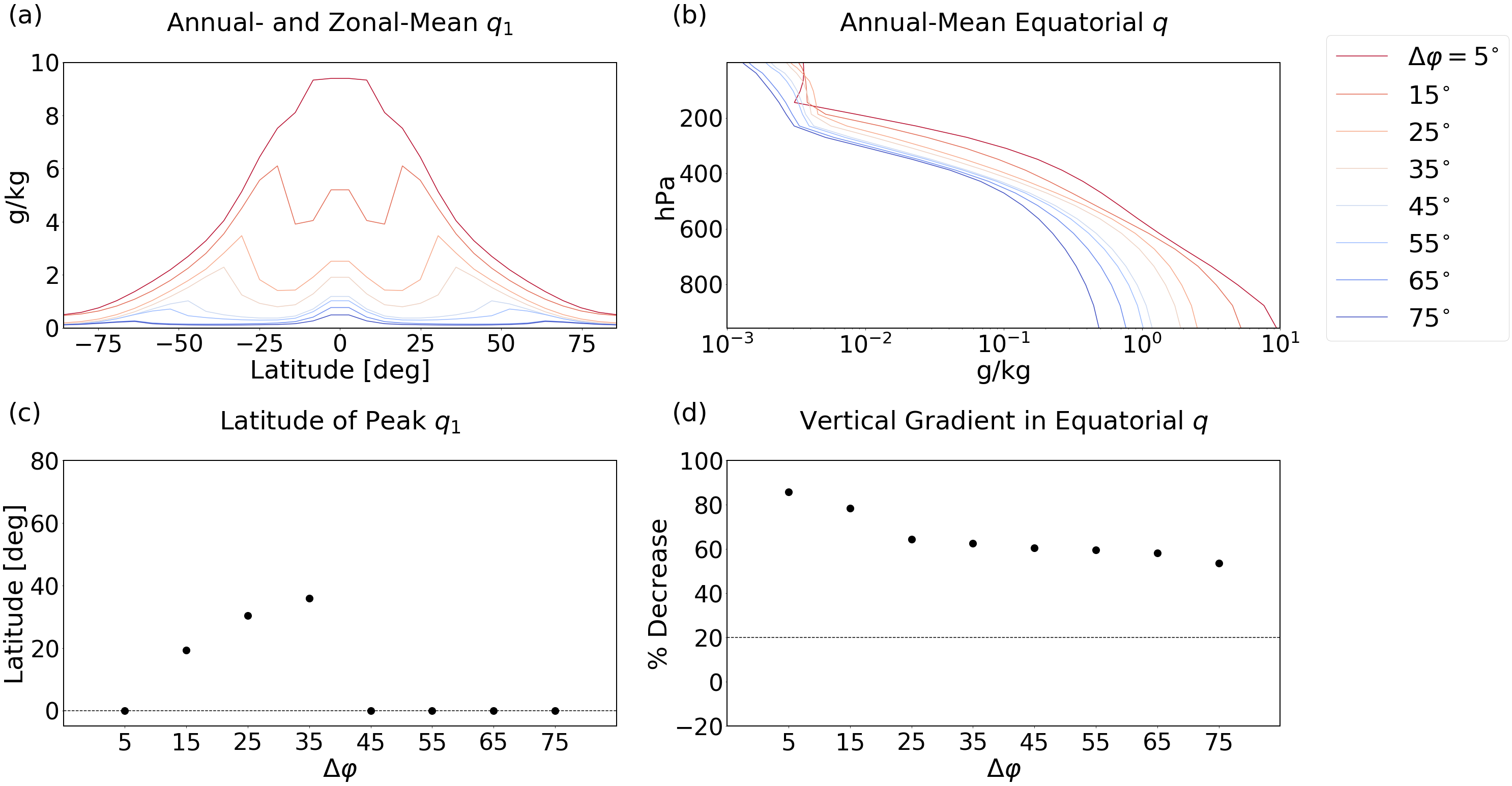}
\caption{Specific humidity analysis for the $\Delta\varphi$ experiments. Figures (a) and (b) are equivalent to Figures \ref{re_exp_sh}(a) and (b), respectively, for these experiments. Figures (c) and (d) are equivalent to Figures \ref{re_exp_sh}(c) and (d), respectively.}
\label{lat_exp_sh}
\end{figure}

We assess the first two Titan-like criteria, OffEq and ConVQ, in Figure \ref{lat_exp_sh}. We find that the presence of the land strip lowers the equatorial value of $q_1$ significantly for experiments with $\Delta\varphi>5^{\circ}$ and $\leq35^{\circ}$. Figure \ref{lat_exp_sh}(c) shows the latitude of peak $q_1$ moving outward from the equator with larger $\Delta\varphi$ until $\Delta\varphi=45^{\circ}$, after which it returns to the equator. The equatorial value appears to decrease less quickly than the midlatitude value for higher $\Delta\varphi$ in \ref{lat_exp_sh}(a), which allows it to reclaim the overall peak. The midlatitude peaks in specific humidity are similar to the values at those latitudes in cases with smaller $\Delta\varphi$. This means they are dependant on the surface temperature at the shorelines, which decreases as $\Delta\varphi$ increases and the shorelines move further poleward. This effect becomes more important than the effect of the land strip on the equatorial peak in specific humidity for $\Delta\varphi\geq45^{\circ}$. We hypothesize that the equatorial peak is primarily dependant on the HC's access to midlatitude moisture, which would be largely cut off once $\Delta\varphi$ exceeds $35^{\circ}$. In contrast, the midlatitude peak's dependence on local surface temperature would continue to be relevant for values of $\Delta\varphi$ well past $45^{\circ}$, allowing for the observed switch in which peak dominates.

Figure \ref{lat_exp_sh}(b) shows the vertical profile of specific humidity at the equator while \ref{lat_exp_sh}(d) shows the percent decrease from the surface to 600hPa pressure level as in Figure \ref{re_exp_sh}(c), d. None of the experiments develop the near-vertical profile of the highest $r_E$ cases, however there is a general trend toward deeper, more uniform vertical profiles with increasing $\Delta\varphi$. 

\begin{figure}[h]
\hspace{0.25\textwidth}\includegraphics[width=0.5\textwidth]{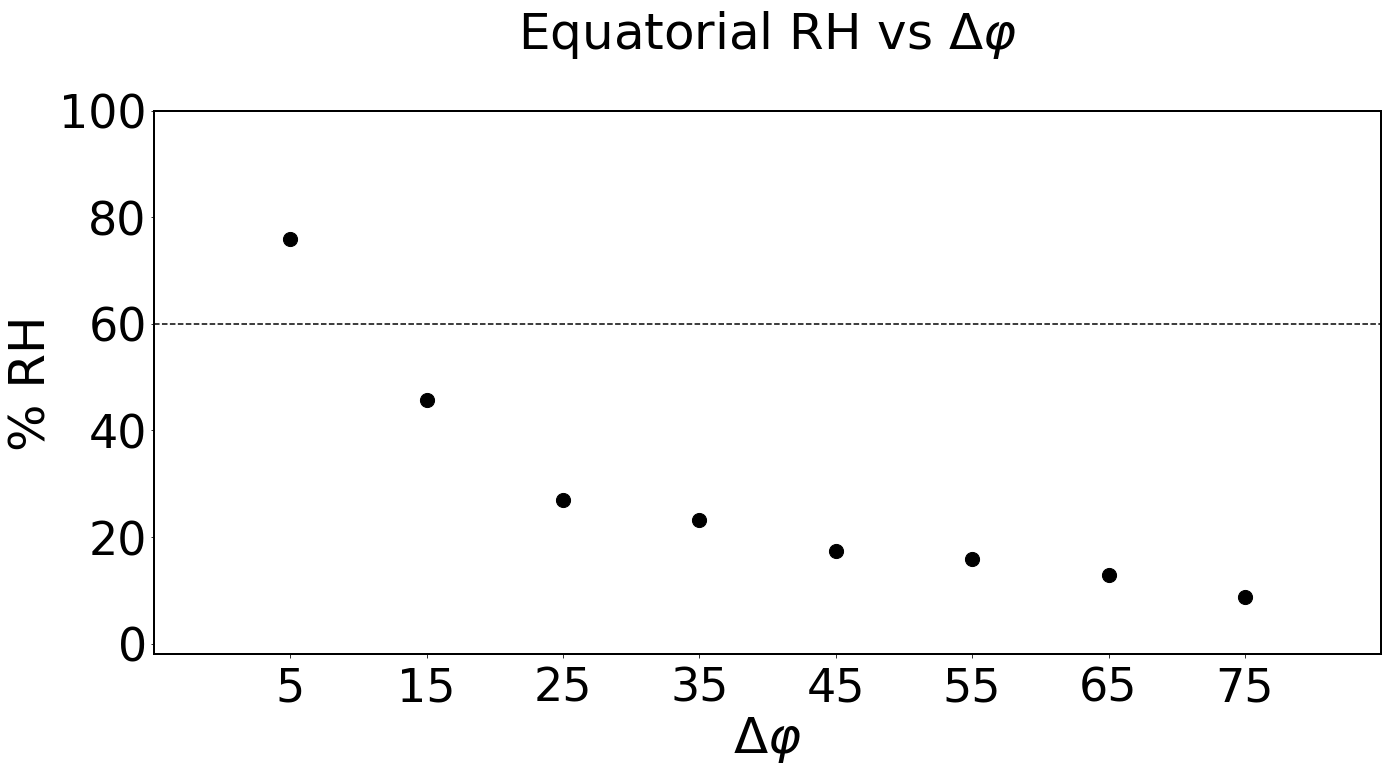}
\caption{RH analysis for the variable-$\Delta\varphi$ experiments, equivalent to Figure \ref{re_exp_rh}.}
\label{lat_exp_rh}
\end{figure}

Despite having no experiments meet the ConVQ criterion, the $\Delta\varphi$ experiments almost all meet the LowRH criterion. Figure \ref{lat_exp_rh} shows the near-surface equatorial- and time-mean RH as a function of $\Delta\varphi$. The RH quickly drops below the 60\% threshold once $\Delta\varphi$ exceeds 5$^{\circ}$, and dips below 20\% for $\Delta\varphi\geq45^{\circ}$. The jump between $\Delta\varphi=5^{\circ}$ and $15^{\circ}$ certainly moves the hydroclimate to the drier equatorial hydrology expected for Titan-like climates.

The results of the $\Delta\varphi$ experiments are summarized in Table \ref{lat_table}. As with the $r_E$ experiments, the LowRH criterion is met by the most experiments, but in contrast the OffEq criterion is met by some experiments while none meet the ConVQ criterion. We conclude that the mixture of dry and wet surfaces has a different impact on the equatorial climate state than a uniform evaporative resistance. In addition, the most Titan-like criteria are only met by experiments with moderately-sized land strips, suggesting there are competing effects of $\Delta\varphi$ and simply increasing it does not produce a fully Titan-like state. We expand on this set of experiments by including two extra parameters, rotation period and condensable volatility. We vary these for a subset of the land strips ($\Delta\varphi=25^{\circ}$, $35^{\circ}$, and $45^{\circ}$), moving from Earth-like values to Titan-like ones.

\begin{table}[h]
\begin{center}
\begin{tabular}{|c|c|c|c|}
\hline
$ $ & \multicolumn{3}{|c|}{Criteria}\\
\cline{2-4}
$ $ & OffEq & ConVQ & LowRH \\
\hline
$\Delta\varphi=5^{\circ}$ & $ $ & $ $ & $ $\\
\hline
$15^{\circ}$ & X & $ $ & X\\
\hline
$25^{\circ}$ & X & $ $ & X\\
\hline
$35^{\circ}$ & X & $ $ & X\\
\hline
$45^{\circ}$ & $ $ & $ $ & X\\
\hline
$55^{\circ}$ & $ $ & $ $ & X\\
\hline
$65^{\circ}$ & $ $ & $ $ & X\\
\hline
$75^{\circ}$ & $ $ & $ $ & X\\
\botline
\end{tabular}
\caption{Criteria matched by each $\Delta\varphi$ experiment.}
\label{lat_table}
\end{center}
\end{table}

\subsection{Variable $T_r$ Experiments}

\begin{figure}[h]

\includegraphics[width=1\textwidth]{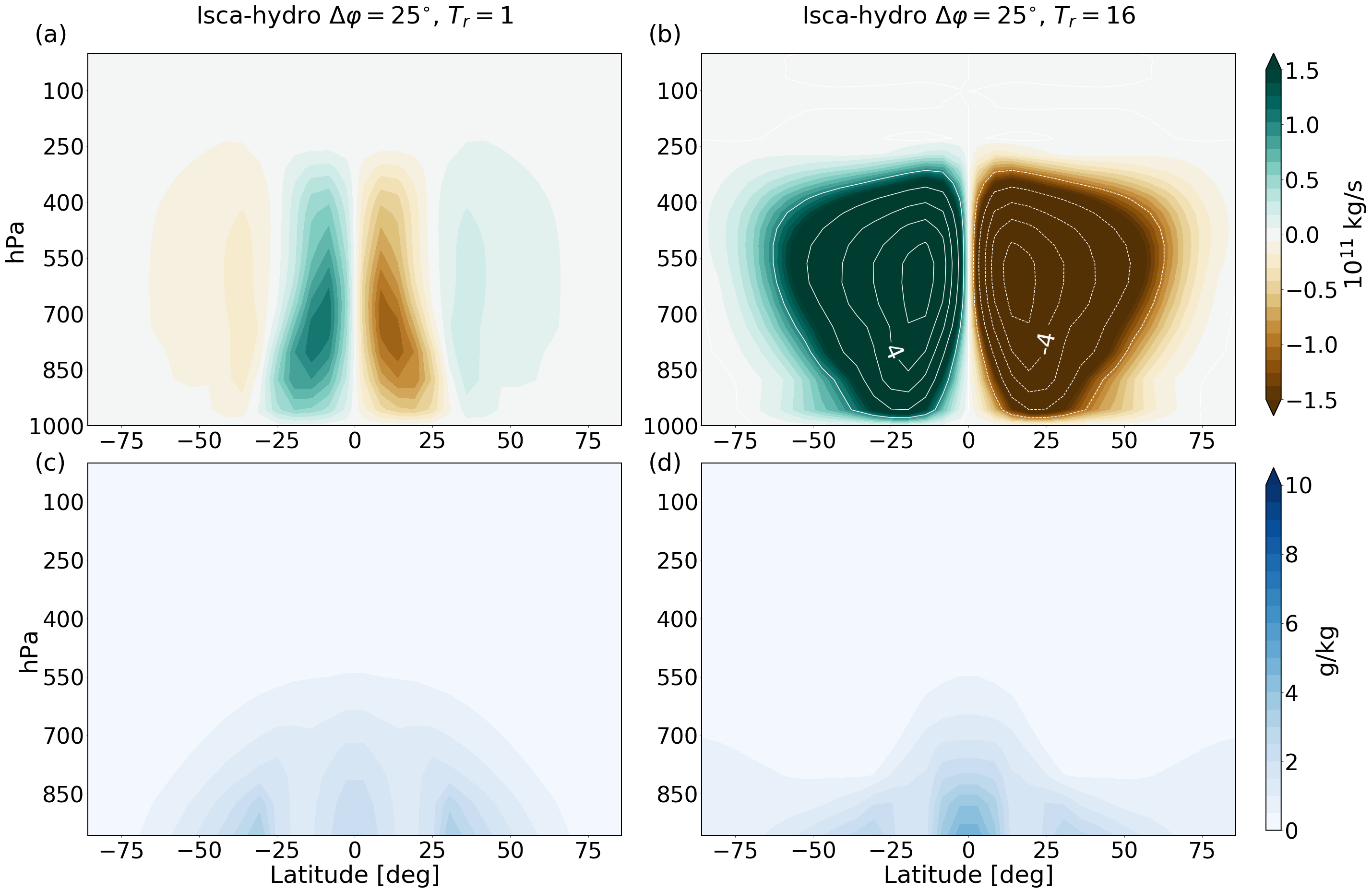}
\caption{Top row: Zonal- and time-mean stream function for the Isca-hydro $\Delta\varphi=25^{\circ}$ experiments with (a) $T_r=1$ and (b) $T_r=16$. We add white contour lines to (b) to show values beyond the range of the filled contours. These contour lines show values in $10^{11}$ kg/s and are spaced $10^{11}$ kg/s apart. Bottom row: Equivalent to the top row but for the specific humidity fields of each experiment. We find a significant increase in HC strength with larger $T_r$ and a corresponding increase in equatorial specific humidity.}
\label{rot_exp_basic_comp}
\end{figure}

Two significant atmospheric circulation patterns affecting the water cycle -- the width of the Hadley circulation and the location of the baroclinic zone -- depend on $T_r$ \citep{Kaspi2015}. We then expect that varying the rotation period will influence the equilibrium distribution of water vapor by influencing the origin of moisture transport relative to shorelines. In the next set of experiments, we vary the rotation period across the values $T_r=1$, 2, 4, 8, and 16-days for three land-strip widths of $\Delta\varphi=25^{\circ}$, $35^{\circ}$, and $45^{\circ}$, and holding all other parameters fixed. As with previous experiments sets, we begin with a look at two end-member cases to assess the broad affects of rotation. Figure \ref{rot_exp_basic_comp}(a) (b) shows the zonal- and time-mean stream function for the experiment with $\Delta\varphi=25^{\circ}$ and $T_r=1$ ($T_r=16$). The intensity of the HC is much greater in the $T_r=16$, as is the equatorial specific humidity. The high equatorial humidity suggests that the higher $T_r$ is allowing the equator greater access to midlatitude moisture by expanding the HC. We will keep this in mind as we analyze the three Titan-like criteria.

\begin{figure}[h]
\includegraphics[width=1\textwidth]{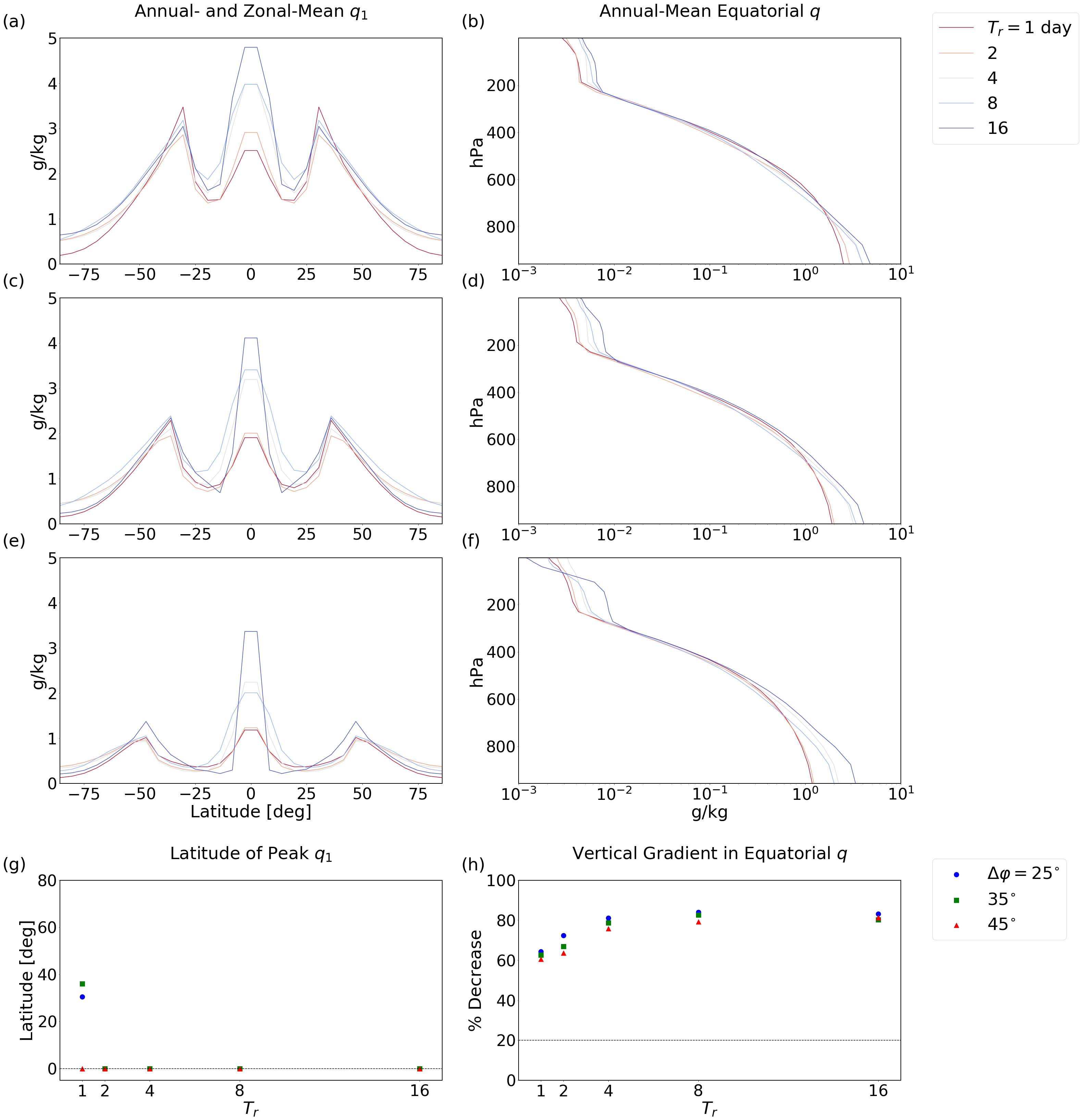}
\caption{Specific humidity analysis for the $T_r$ experiments. Figures (a), (c), and (e) show the zonal- and time-mean $q_1$, as in Figure \ref{lat_exp_sh}(a). Figures (b), (d), and (f) show the time-mean equatorial vertical profile of specific humidity, as in Figure \ref{lat_exp_sh}(b). First row is for the experiments with $\Delta\varphi=25^{\circ}$, second row for $35^{\circ}$, and third row $45^{\circ}$. Figure (g) shows the latitude of peak $q_1$ as a function of $T_r$ for each value of $\Delta\varphi$, while (h) shows the percent decrease in specific humidity between the surface and 600hPa pressure level as a function of $T_r$.}
\label{rot_exp_sh_vsh}
\end{figure}

Figures \ref{rot_exp_sh_vsh}(a), c, and e show zonal- and time-mean $q_1$ for the $T_r$ experiments. Only two experiments meet the OffEq criterion, the $\Delta\varphi=25^{\circ}$ and $35^{\circ}$ cases where $T_r=1$ day. The equatorial peaks in $q_1$ appear to increase with higher $T_r$, while the midlatitude peaks are constant or slightly decreasing. This allows the equatorial peak to dominate in all experiments with $T_r\geq2$. We hypothesize this is due to the width of the Hadley circulation in relation to the width of the land strip. In the 1-day cases, the HC extent is more likely to be contained by the land strip, cutting it off from midlatitude moisture and allowing it to accumulate in the midlatitudes. 


\begin{figure}[h]
\hspace{0.1\textwidth}\includegraphics[width=0.8\textwidth]{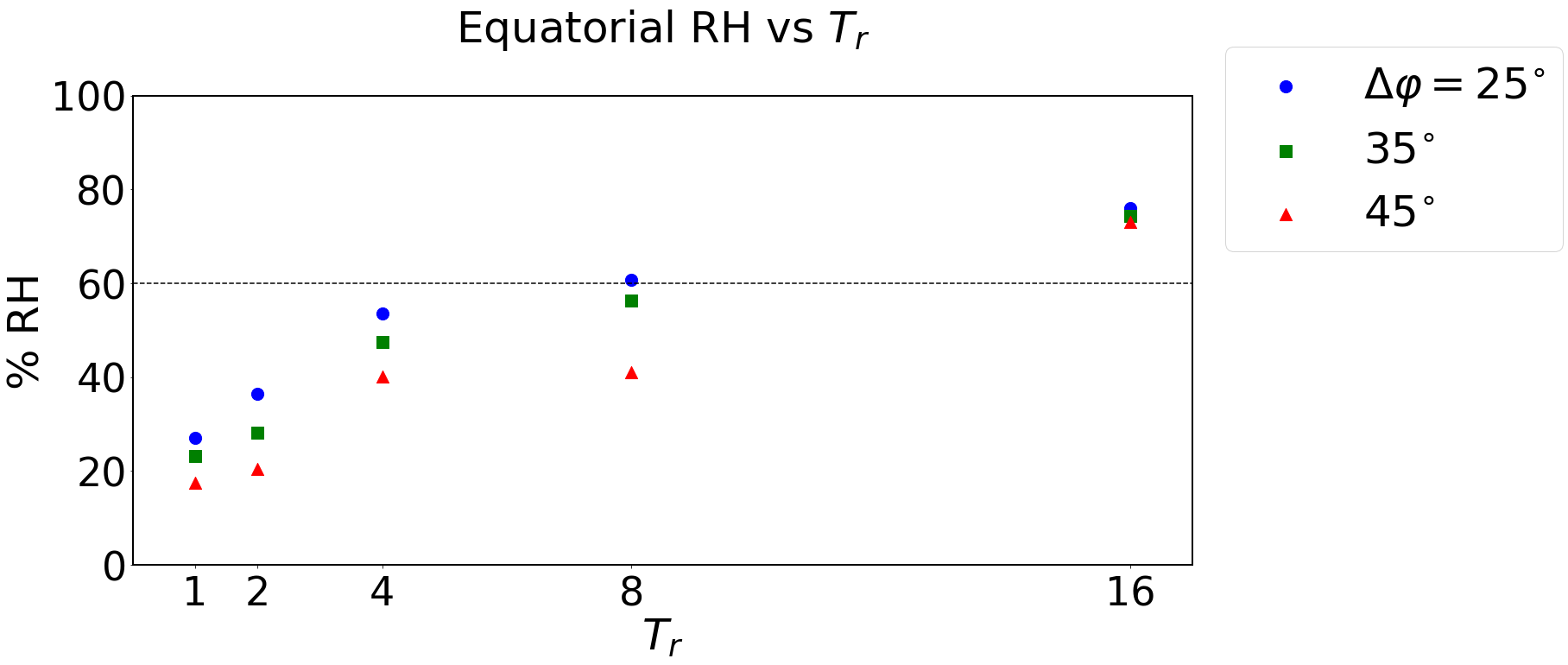}
\caption{RH analysis for the $T_r$ experiments, equivalent to Figure \ref{re_exp_rh}.}
\label{rot_exp_rh}
\end{figure}

Similar to the $\Delta\varphi$ experiments, no $T_r$ experiments meet the ConVQ criterion. There is in fact a trend away from the threshold (Figure \ref{rot_exp_sh_vsh}(h)) with higher $T_r$, despite Titan itself having $T_r=16$. In fact it is possible that for values of $T_r<1$ an experiment could meet the ConVQ criterion. Reconciling this with Titan's high value of $T_r$ brings us back to the land strip itself: the relevant determiner for the ConVQ criterion is the relation between the HC and land strip, rather than either in isolation. The effect of a wider land strip (increasing $\Delta\varphi$, Figure \ref{lat_exp_sh}(d)) should be equivalent to the effect of a narrower HC (decreasing $T_r$, Figure \ref{rot_exp_sh_vsh}(h)). This means that even though Titan has a large $T_r$, it is made up for by an also large $\Delta\varphi$ which dominates the overall hydroclimate.

Figure \ref{rot_exp_rh} shows that all but one experiment with $T_r\leq8$ meet the LowRH criterion, with that one experiment, $\Delta\varphi=25^{\circ}$ and $T_r=8$, narrowly missing. All experiments with $T_r=16$ fail to meet the LowRH criterion, and all have very similar values. In addition, their RH values are close to those of the $\Delta\varphi=5^{\circ}$ experiment, suggesting their equatorial regions have access to abundant moisture despite the land strips.

Overall, experiments with smaller $T_r$ and larger $\Delta\varphi$ were more likely to meet the three criteria. The latter is within expectations, as Titan has a dry tropics and subtropics which mirrors our land strips. The former, though, is unexpected in that Titan is a slow rotator, $T_r=16$, but in our experiments slow rotation moved the equatorial climate \emph{away} from being Titan-like. We discuss this further in Section 5.

\begin{table}[h]
\begin{center}
\begin{tabular}{|c|c|c|c|c|}
\hline
\multicolumn{2}{|c|}{$ $} & \multicolumn{3}{|c|}{Criteria}\\
\hline
$\Delta\varphi$ & $T_r$ & OffEq & ConVQ & LowRH \\
\hline
$ $ & 1-day & X & $ $ & X\\
\cline{2-5}
$ $ & 2 & $ $ & $ $ & X\\
\cline{2-5}
$25^{\circ}$ & 4 & $ $ & $ $ & X\\
\cline{2-5}
$ $ & 8 & $ $ & $ $ & $ $\\
\cline{2-5}
$ $ & 16 & $ $ & $ $ & $ $\\
\hline
$ $ & 1-day & X & $ $ & X\\
\cline{2-5}
$ $ & 2 & $ $ & $ $ & X\\
\cline{2-5}
$35^{\circ}$ & 4 & $ $ & $ $ & X\\
\cline{2-5}
$ $ & 8 & $ $ & $ $ & X\\
\cline{2-5}
$ $ & 16 & $ $ & $ $ & $ $\\
\hline
$ $ & 1-day & $ $ & $ $ & X\\
\cline{2-5}
$ $ & 2 & $ $ & $ $ & X\\
\cline{2-5}
$45^{\circ}$ & 4 & $ $ & $ $ & X\\
\cline{2-5}
$ $ & 8 & $ $ & $ $ & X\\
\cline{2-5}
$ $ & 16 & $ $ & $ $ & $ $\\
\botline
\end{tabular}
\caption{Criteria matched by each $T_r$ experiment.}
\label{rot_table}
\end{center}
\end{table}

\subsection{Variable $\xi$ Experiments}

\begin{figure}[h]

\includegraphics[width=1\textwidth]{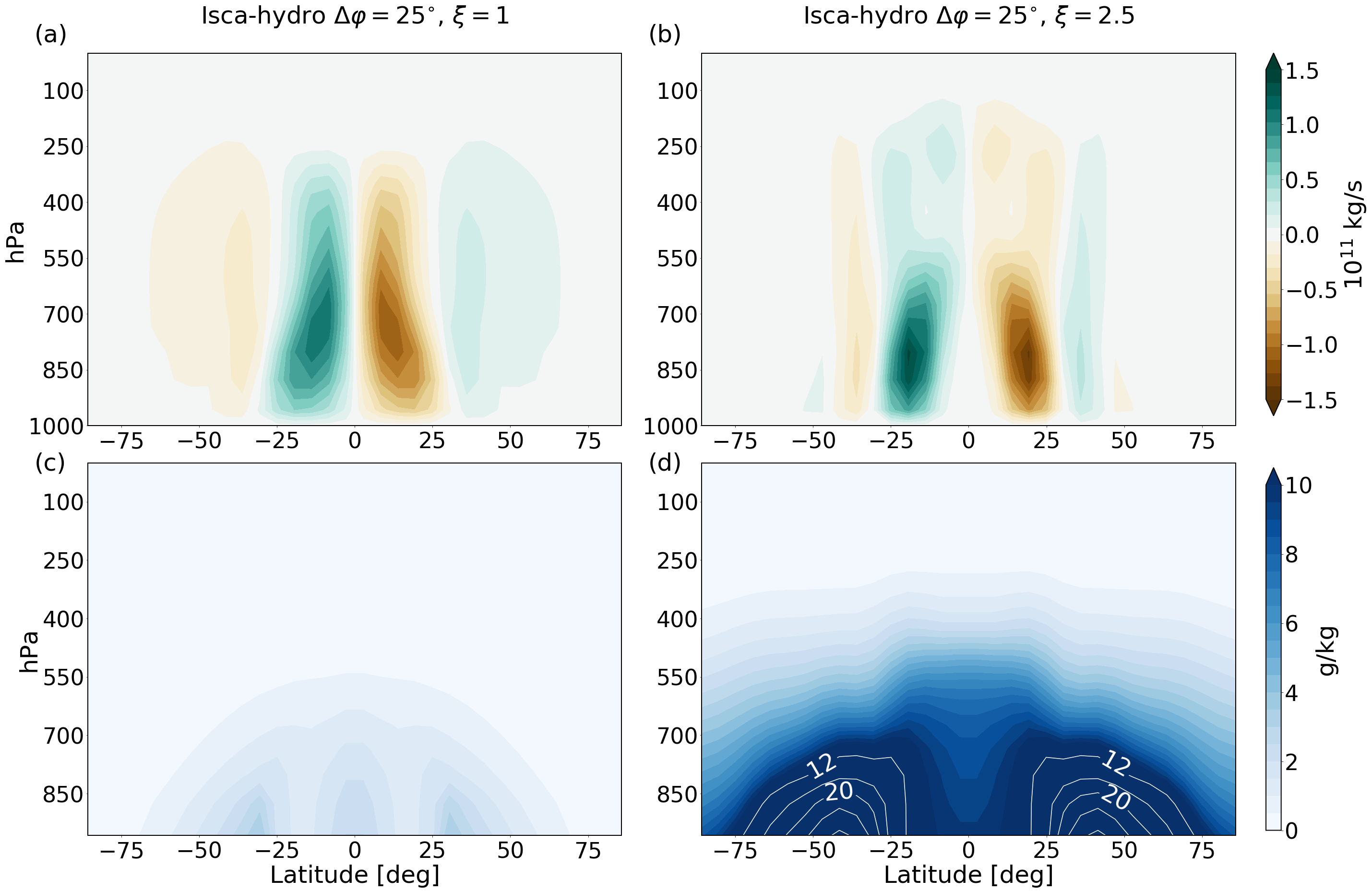}
\caption{Top row: Zonal- and time-mean stream function for the Isca-hydro $\Delta\varphi=25^{\circ}$ experiments with (a) $\xi=1$ and (b) $\xi=2.5$. Bottom row: Equivalent to the top row but for the specific humidity fields of each experiment. We add white contours in (d) to show the specific humidity structure. These contours show values in g/kg and are spaced 4 g/kg apart. We find that the HC both contracts vertically and moves outward from the equator with higher $\xi$.}
\label{xi_exp_basic_comp}
\end{figure}

In the final set of experiments, we vary the water volatility parameter, $\xi$, to mimic the higher volatility of methane on Titan. Based on in-situ measurements of Titan's atmosphere by the Huygens probe \citep{niemann2005abundances}, we can approximate the surface mole fraction of methane to N$_2$ on Titan as 0.05 and the RH as 50\%. We can convert this to a specific humidity value by multiplying the ratio of the molecular masses of methane and N$_2$,
\begin{linenomath*}
\begin{equation}
q_{Titan}\approx0.05\times\frac{16}{28}\approx 30 \textrm{g/kg}
\end{equation}
\end{linenomath*}

\noindent If we take $25^{\circ}$C (298.15K) as an arbitrary but representative value for Earth's surface temperature, we can apply the Clausius-Clapeyron equation to get a representative saturation vapor pressure $e_s$, which in turn can be converted into a saturation specific humidity, $q_s$,
\begin{linenomath*}
\begin{eqnarray}
e_s & = & e_0\exp\left[\frac{L_v}{R_v}\left(\frac{1}{T_0}-\frac{1}{298.15}\right)\right]\\
q_s & = & \frac{e_sR_d}{p_sR_v}
\end{eqnarray}
\end{linenomath*}

\noindent where $e_0=611$ hPa, $L_v=2.25\times10^6$ J/kg, $R_v=462$ J/kg/K, $T_0=273.15$ K, $R_d=287$ J/kg/K, and $p_s=1000$ hPa. This yields $q_s\approx20$ g/kg, so if we assume 50\% RH as in the Titan observations we are left with $q_{Earth}\approx10$ g/kg for our Earth value of specific humidity. This means we can calculate an effective $\xi$ value for Titan, $\xi_{eff}$,
\begin{linenomath*}
\begin{equation}
\xi_{eff}=\frac{q_{Titan}}{q_{Earth}} \approx 3
\end{equation}
\end{linenomath*}

\noindent This is slightly above the maximum $\xi$ value of 2.5 used in this work, but this is a highly approximate value and based on the following results this difference does not have a significant impact on our analysis.

We use the $\Delta\varphi=25^{\circ}$ experiments with $\xi=1$ and $\xi=2.5$ as the end-members for comparison in Figure \ref{xi_exp_basic_comp}. Figures \ref{xi_exp_basic_comp}(a) and b are the zonal- and time-mean stream functions for each case, while c and d are the equivalent specific humidity fields. Unlike in previous comparisons, we find that the vertical contraction of the HC in \ref{xi_exp_basic_comp}(b) corresponds to an increase in specific humidity in \ref{xi_exp_basic_comp}(d). In fact, the specific humidity is far higher than any previous case, as might be expected with high values of $\xi$. The effect on the HC may instead be connected to the RH, which does not increase in line with specific humidity. Another interesting feature is the shift in the HC peak away from the equator with higher $\xi$. The peak latitude moves from roughly $10^{\circ}$ latitude to $20^{\circ}$, with the lower part of the circulation moving further out than the upper. There is also a suggestion of an equatorial reversal near the surface.

\begin{figure}[h]


\includegraphics[width=1\textwidth]{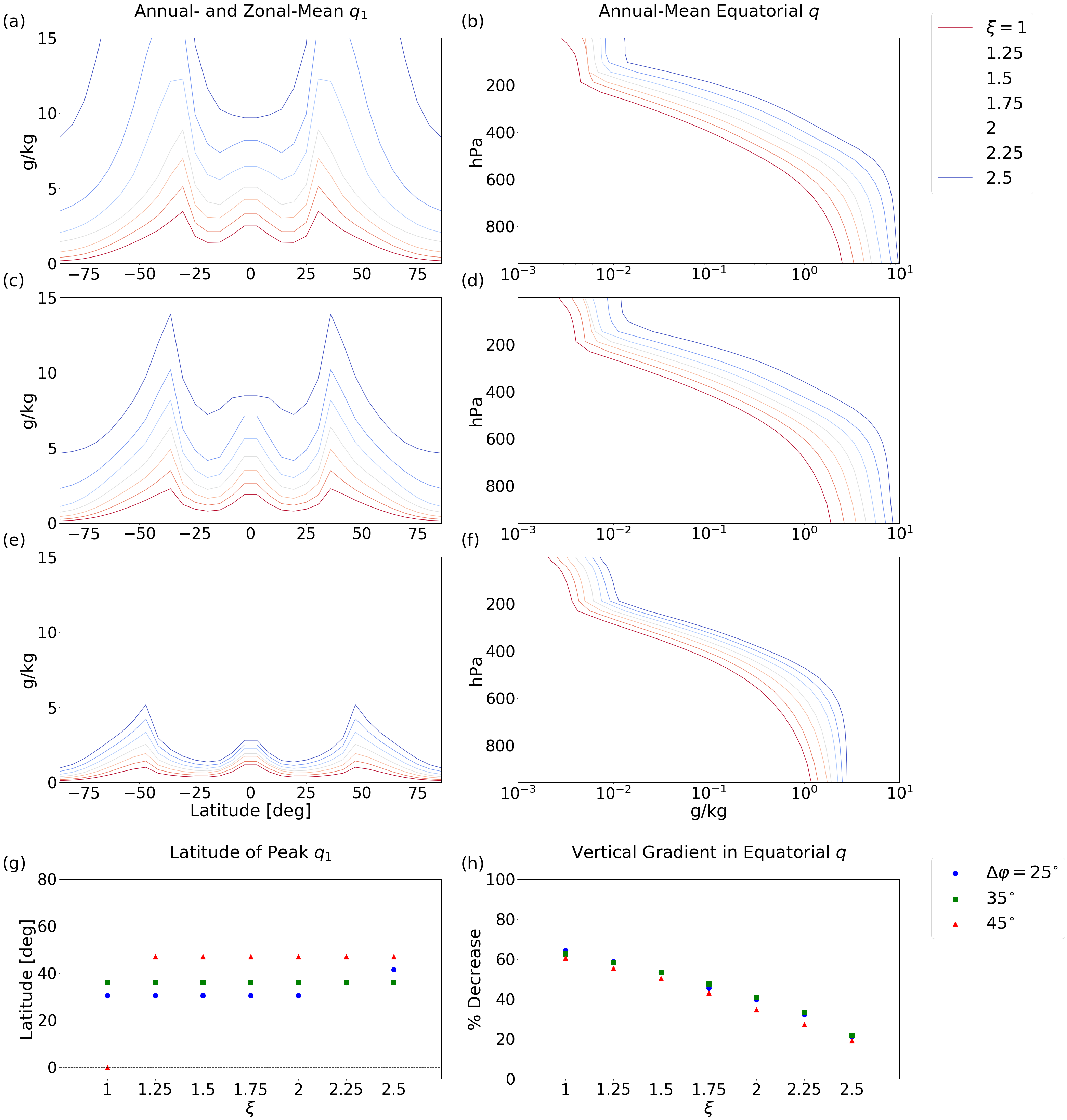}
\caption{Left column: Equivalent to Figures \ref{rot_exp_sh_vsh}(a), (c), (e), and (g) but for the $\xi$ experiments. Right column: Equivalent to Figures \ref{rot_exp_sh_vsh}(b), (d), (f), and (h).}
\label{xi_exp_sh_vsh}
\end{figure}

Figures \ref{xi_exp_sh_vsh}(a), c, and e show $q_1$ for each $\xi$ experiment. All cases except for one have their humidity maximum off of the equator (\ref{xi_exp_sh_vsh}(g)), although for most the equator is a local maximum. $q_1$ increases everywhere with higher $\xi$, although this effect is less significant at the equator compared to the midlatitudes. The increased trend is expected given that higher $\xi$ represents an increase in the amount of water vapor needed to saturate air, while the smaller effect at the equator is likely due to the land strips limiting access to moisture. We conclude that all but one case ($\Delta\varphi=45^{\circ}$, $\xi=1$) meet the OffEq criterion.


Figures \ref{xi_exp_sh_vsh}(b), d, and f show the vertical specific humidity profiles at the equator. Multiple experiments have nearly vertical profiles in the lower atmosphere, and three just reach the $20\%$ gradient threshold (\ref{xi_exp_sh_vsh}(h)). These three experiments all have $\xi=2.5$, demonstrating an inverse relationship between meeting the ConVQ criterion and the magnitude of specific humidity when compared to the aquaplanet experiments, i.e. the experiments with the highest humidities are the ones with constant vertical profiles. While seemingly contradictory, Titan itself has a more volatile condensable in methane than water on Earth, and a correspondingly higher specific humidity of methane in its atmosphere despite being ``drier'' in RH terms.

\begin{figure}[h]
\includegraphics[width=1\textwidth]{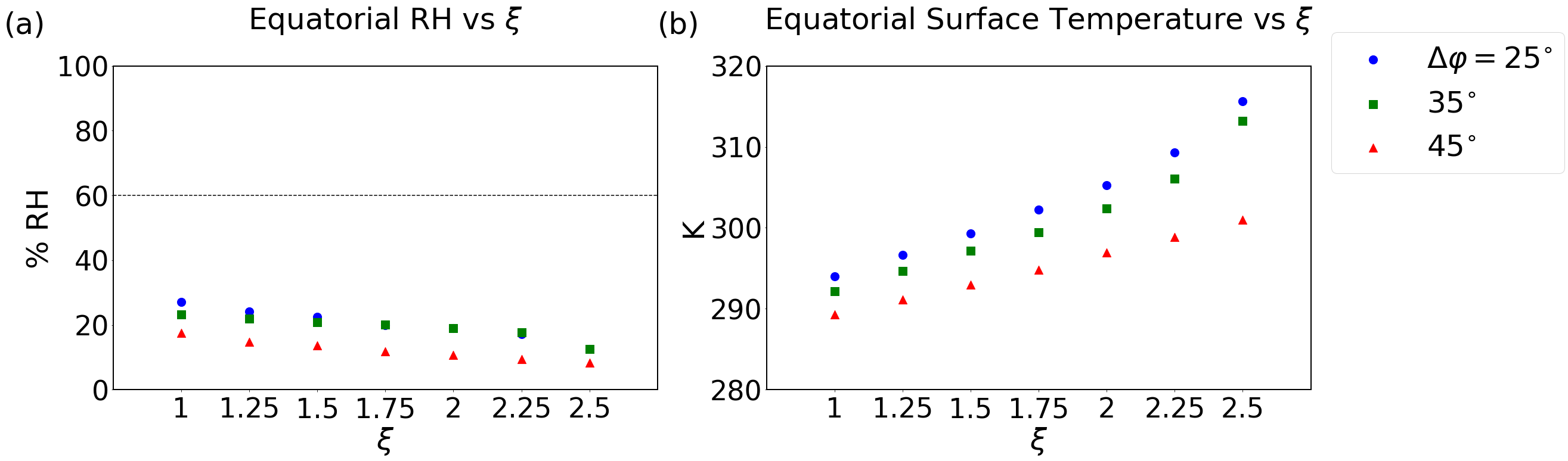}
\caption{(a): RH analysis for the $\xi$ experiments, equivalent to Figure \ref{re_exp_rh}. (b): equivalent to (a) but for surface temperature.}
\label{xi_exp_rh}
\end{figure}

Figure \ref{xi_exp_rh}(a) shows the time-mean equatorial near-surface RH. All cases are well below the 60\% threshold to meet the LowRH criterion. There is a weak trend towards lower RH with higher $\xi$. The fact that the higher $\xi$ experiments have both low RH and high $q_1$ means that the temperature, and thus the $q_s^*$, are more strongly increasing with $\xi$ than $q_1$. We can see the increase in equatorial surface temperature in Figure \ref{xi_exp_rh}(b) as a function of $\xi$. Even in the cooler $\Delta\varphi=45^{\circ}$ experiments there is an increase of over 10K between $\xi=1$ and $\xi=2.5$. The net effect is a ``drier'' (lower RH) climate in the high-$\xi$ cases despite having correspondingly higher $q_1$.

We find that only the cases with $\xi=2.5$ meet the ConVQ criterion, while all but one of the $\xi$ experiments meet the OffEq and LowRH criteria. We conclude that the primary effect of $\xi$ is on the ConVQ criterion, with the OffEq and LowRH criteria influenced more by the presence of the land strips. The full results of the $\xi$ experiments are summarized in Table \ref{xi_table}.

\begin{table}[h]
\begin{center}
\begin{tabular}{|c|c|c|c|c|}
\hline
\multicolumn{2}{|c|}{$ $} & \multicolumn{3}{|c|}{Criteria}\\
\hline
$\Delta\varphi$ & $\xi$ & OffEq & ConVQ & LowRH \\
\hline
$ $ & 1.00 & X & $ $ & X\\
\cline{2-5}
$ $ & 1.25 & X & $ $ & X\\
\cline{2-5}
$ $ & 1.50 & X & $ $ & X\\
\cline{2-5}
$25^{\circ}$ & 1.75 & X & $ $ & X\\
\cline{2-5}
$ $ & 2.00 & X & $ $ & X\\
\cline{2-5}
$ $ & 2.25 & X & $ $ & X\\
\cline{2-5}
$ $ & 2.50 & X & X & X\\
\hline
$ $ & 1.00 & X & $ $ & X\\
\cline{2-5}
$ $ & 1.25 & X & $ $ & X\\
\cline{2-5}
$ $ & 1.50 & X & $ $ & X\\
\cline{2-5}
$35^{\circ}$ & 1.75 & X & $ $ & X\\
\cline{2-5}
$ $ & 2.00 & X & $ $ & X\\
\cline{2-5}
$ $ & 2.25 & X & $ $ & X\\
\cline{2-5}
$ $ & 2.50 & X & X & X\\
\hline
$ $ & 1.00  & $ $ & $ $ & X\\
\cline{2-5}
$ $ & 1.25 & X & $ $ & X\\
\cline{2-5}
$ $ & 1.50 & X & $ $ & X\\
\cline{2-5}
$45^{\circ}$ & 1.75 & X & $ $ & X\\
\cline{2-5}
$ $ & 2.00  & X & $ $ & X\\
\cline{2-5}
$ $ & 2.25 & X & $ $ & X\\
\cline{2-5}
$ $ & 2.50 & X & X & X\\
\botline
\end{tabular}
\caption{Criteria matched by each $\xi$ experiment.}
\label{xi_table}
\end{center}
\end{table}

\section{Discussion}

Overall, the parameter most effective at achieving the three Titan-like criteria was $\xi$. The only experiments to meet all three criteria had $\xi=2.5$, the highest value we used for the parameter. The other parameters allowed experiments to meet two of the three criteria, but never had any meet all three. The $r_E$ experiments met more criteria with higher $r_E$, but were unable to meet the OffEq criterion due to $r_E$ being applied uniformly to the entire surface. The $\Delta\varphi$ experiments met two criteria for $\Delta\varphi=15^{\circ}$, $25^{\circ}$, and $35^{\circ}$, but only one for higher values of $\Delta\varphi$. This makes it hard to define a clear correlation between $\Delta\varphi$ and our Titan-like criteria. As $\Delta\varphi$ increases, the peaks in $q_1$ associated with the continent's shorelines decrease (Figure \ref{lat_exp_sh}(a)). This decrease roughly follows the curve of the $\Delta\varphi=5^{\circ}$ case, from which we infer the peaks represent the temperature at the shoreline and closely match the value at that latitude in experiments with narrower land strips. For land strips with shorelines well into the midlatitudes the local temperature is cold enough to suppress the local peak in $q_1$ to the point where it is no longer larger than the equatorial value. The equatorial values change more slowly than the shoreline values for $\Delta\varphi\geq35^{\circ}$, perhaps reflecting the HC being cutoff from the ocean (Figure \ref{hcwidth}(a)).

The final parameter, $T_r$, also had experiments in its set meet at most two criteria. The experiments that met two criteria all had $T_r=1$, while experiments with $T_r=16$ met none of the criteria. This means that smaller $T_r$ correlates with meeting more Titan-like criteria, despite Titan itself having $T_r=16$. It is then, perhaps, surprising that its equator is able to maintain such a dry climate. Based on our experiments, the volatility of the surface condensable may be the primary factor in setting Titan's low-$q_1$ equatorial climate. The highest-$\xi$ experiments all met the three criteria, and were the only ones to do so. Expanding the experimental range to include experiments with simultaneously high $\xi$ and $T_r$ will allow us to better understand Titan's climate, which we leave to future work.

\subsection{Moisture Transport Analysis}

\begin{figure}[h]
\includegraphics[width=1\textwidth]{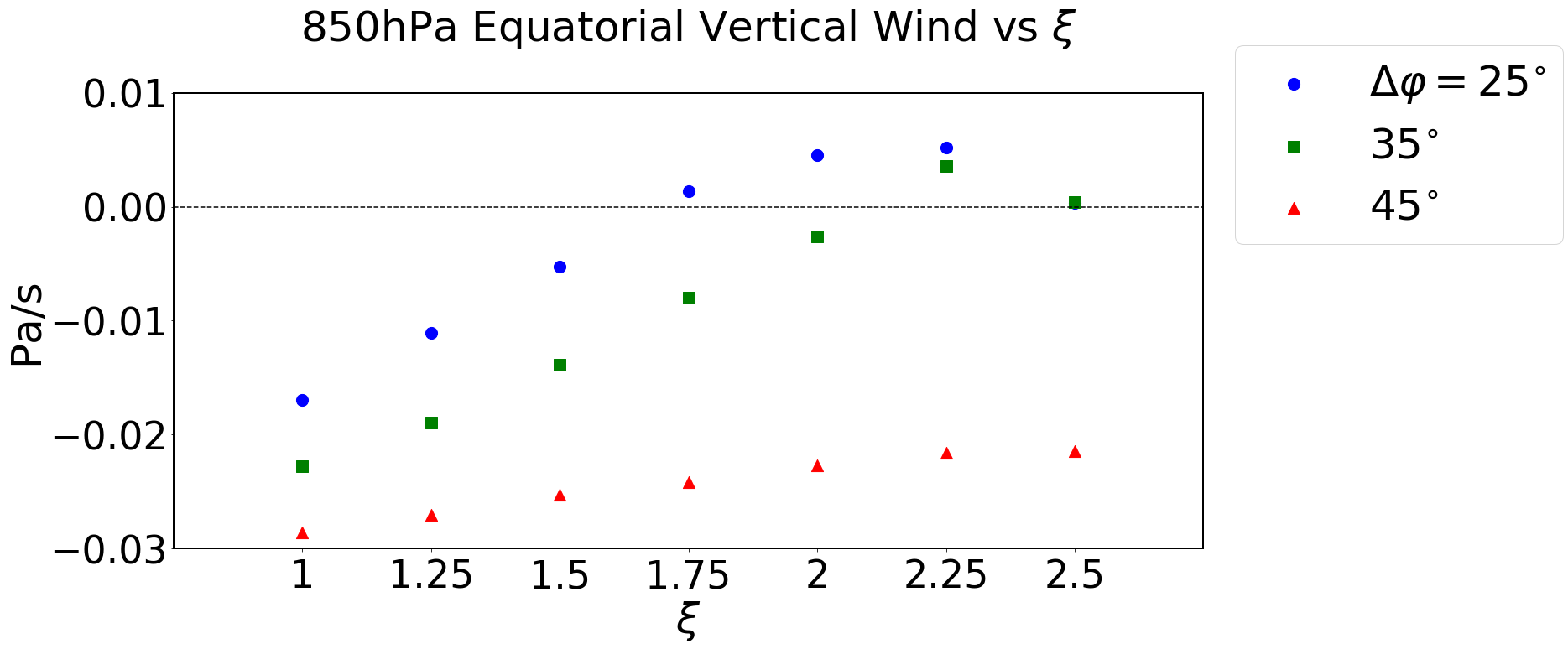}
\caption{Zonal- and time-mean vertical winds at the 850hPa pressure level for the $\xi$ experiments. Units are in Pa/s, so positive values (above dotted line) mean downward motion.}
\label{wcomp}
\end{figure}

We have identified three experiments that meet all of our Titan-like criteria, but do not know how closely they match Titan-like dynamics. Titan's equatorial humidity is hypothesized to be set either by virga and vertical advection downward from the upper troposphere or slantwise, horizontal-vertical moisture transport by baroclinic eddies. It is unlikely to be the result of surface evaporation with low-level moisture advection. In order to identify the source of equatorial moisture in our experiments, we first look at the vertical wind above the equator. Figure \ref{wcomp} shows the time-mean vertical wind at the 850hPa pressure level over the equator for the $\xi$ experiments. We find that the upward winds weaken with higher $\xi$, and even shift direction for the experiments with $\Delta\varphi=25^{\circ}$ and $35^{\circ}$ at high enough values of $\xi$. There is a subsequent weakening of these downward winds between the $\xi=2.25$ and 2.5 cases for both $\Delta\varphi$ sets, meaning the relation between $\xi$ and vertical wind velocity is non-monotonic for the narrower land strips. Considering that the experiments with the strongest downward vertical winds did not meet all three criteria while at least one experiment with upward winds did, we conclude that this alone does not determine whether an experiment will be Titan-like.

\begin{figure}[h]
\includegraphics[width=1\textwidth]{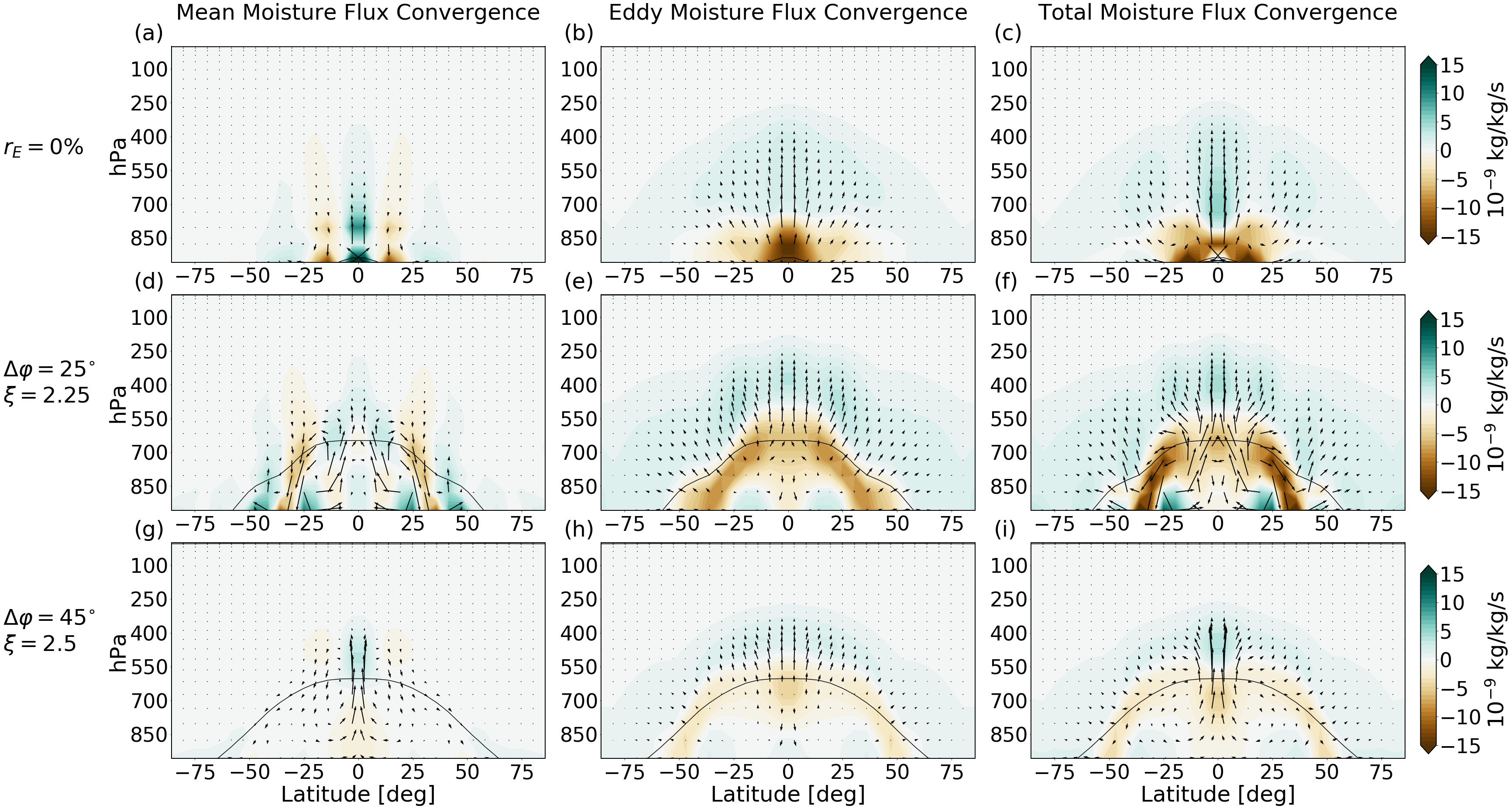}
\caption{Left column: The convergence of the mean moisture flux term with the vector-field overlaid for the Isca-hydro $r_E=0\%$ aquaplanet experiment and two $\xi$ experiments. Center column: The equivalent plots of the eddy term of the moisture flux. Right column: The equivalent plots of the total moisture flux. Additionally overlaid in each plot is the isotherm of last saturation at the equator, equivalent to the equator's mean surface dew point temperature. The top row shows the Isca-hydro $r_E=0\%$ experiment, the middle row the $\Delta\varphi=25^{\circ}$, $\xi=2.25$ experiment, and the bottom row the $\Delta\varphi=45^{\circ}$, $\xi=2.5$ experiment.}
\label{mflux}
\end{figure}

In order to better understand the experiments with upward vertical winds and those that met the three criteria, we take two experiments from each group and plot the zonal- and time-mean moisture convergence with the corresponding moisture transport vectors overlaid (Figure \ref{mflux}). For additional comparison, the Isca-hydro experiment with $r_E=0\%$ is also included in Figure \ref{mflux}. We overlay the atmospheric isotherm at the dew point temperature of the equatorial surface, $T_{d,eq}$. This isotherm is only just above the equatorial surface in the $r_E=0\%$ experiment (\ref{mflux}(a)-c), as might be expected for an aquaplanet with abundant water at the equator. However, for the two $\xi$ experiments this isotherm is well above the surface at the equator and only meets the surface at latitudes greater than $50^{\circ}$. Considering that the widest land strip of these two is only $\Delta\varphi=45^{\circ}$, this strongly suggests that air parcels are not moving into the equator horizontally from the continent's edge.

If moisture is not coming into the equator horizontally, then where is it coming from? Looking at the mean moisture flux vectors in Figures \ref{mflux}(d) and g, there are many arrows showing horizontal movement into the near-surface equator. But these arrows all originate at latitudes within the land strips of their respective experiments. As such, the actual source of the moisture must be elsewhere. In \ref{mflux}(d) the flux vectors in the near-surface equatorial region are downward, in line with the winds shown in Figure \ref{wcomp}. This is also true for the eddy terms in \ref{mflux}(e). This suggests the only possible source of moisture for the equatorial surface is from above, advected in by downwelling air. There is a region of weak moisture flux divergence in the mean term at around the 700hPa level, with weak convergence below. The flux vectors at this level also show primarily horizontal flux into the equator, leading us to conclude that moisture in the $\Delta\varphi=25^{\circ}$, $\xi=2.25$ experiment is first brought to the equator aloft in the mid-troposphere before sinking to the surface.

The above case seems to match one hypothesis for Titan's constant specific humidity profile, specifically that the equatorial surface moisture is sourced from above and supplied by downward moisture flux. It does not, however, match all three Titan-like criteria, in contrast to the case shown in Figures \ref{mflux}(g)-i. We find no such downward flux in the mean terms (\ref{mflux}(g)), and instead there is a fairly straightforward circulation of moisture up at the equator and downward in the subtropics. This belies the fact that $\Delta\varphi$ is well beyond the subtropics in this experiment, meaning the return flow to the equator has no access to oceanic moisture except at the very edges of the land strip. The eddy fluxes appear to compete with the mean, even showing downward flux at the equator despite the lack of downward vertical winds. The mean terms dominate in the flux vectors while the eddy terms dominate in the convergence (\ref{mflux}(i)). Based on the total terms, moisture diverges from a near-surface region close to the edge of the land strip and then flows equatorward. However, this divergence aligns with the dewpoint isotherm slightly above the surface, around 800hPa, where flux vectors are pointing downward as part of the sinking branch of the HC. This could mean that the moisture source for the equator is not the shoreline, but rather an elevated region of the subtropics. This would suggest that the equatorial surface is largely cut off from outside moisture, and its local humidity would match that of the entire HC circulation. We can see evidence of this in \ref{mflux}(i), as the flux vectors flowing out from the equator closely follow the dewpoint isotherm as they descend towards edge of the land strip.

\begin{figure}[h]
\includegraphics[width=1\textwidth]{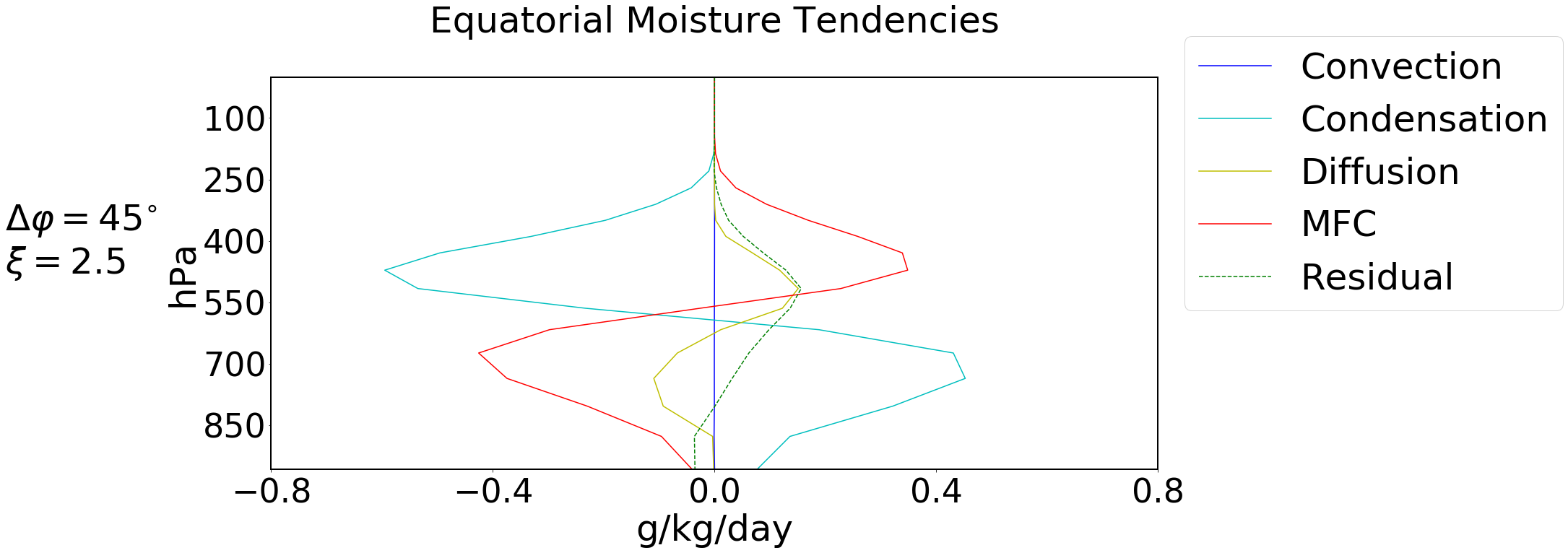}
\caption{Zonal- and time-mean moisture tendencies in the equatorial column for the $\Delta\varphi=45^{\circ}$, $\xi=2.5$ experiment from Figure \ref{mflux}. Positive values mean moisture is being added to the air at that level. The Residual is defined as the value necessary to balance the other four tendencies.}
\label{mtends}
\end{figure}

We further explore the mechanisms of this transport in Figure \ref{mtends}. This figure shows the moisture tendency terms for each vertical level at the equator of the $\Delta\varphi=45^{\circ}$, $\xi=2.5$ experiment. Positive values mean moisture is being added to the atmosphere while negative means moisture is being removed. The first three tendency terms are convection, condensation, and diffusion, which are all directly diagnosed in the model. The fourth term is the moisture flux convergence (MFC), which is the $5^{\circ}$N/S mean of the Total Moisture Flux Convergence from Figure \ref{mflux}. We also add a Residual, which is the value necessary to balance the four tendencies and represents the error of the MFC.

We find that, unlike the Earth-like state, the equatorial convection tendency is zero at all levels -- convection has fully shut off in this experiment. The lower troposphere is primarily characterized by moisture flux divergence (shown as negative MFC in the figure) balanced by re-evaporation (shown as positive condensation), with the roles swapping in the upper troposphere. This means moisture is added to the lower troposphere only via evaporation of rain from the condensation scheme and then removed primarily via MFC and a small amount of diffusion. The opposite then is true of the upper troposphere, where moisture is transported in and then condensed out to produce the rain that supplies the lower levels. The lower atmosphere of this system is only supplied with moisture via re-evaporation, which must necessarily come from above. This may confirm our hypothesis that the equatorial surface is not supplied with moisture via low-level horizontal transport but rather exclusively by downward moisture flux from the upper troposphere. In Figure \ref{mflux}(i) the flux vectors showing horizontal transport into the upper (lower) troposphere at the equator are diverging (converging) in an area of overall convergence (divergence). This means the MFC is primarily due to the vertical component of moisture flux. A consistent interpretation is that falling precipitation completely re-evaporates and the upward transport returns moisture in a closed loop, further supporting our hypothesis that the equatorial water cycle is largely cut off from the high-latitude oceans.

\subsection{Effect of HC Width on Equatorial Moisture}

Our simulations indicate that one way to achieve a Titan-like state on an Earth-like planet may be to have a large equatorial continent ($\Delta\varphi$) and high condensable volatility ($\xi$). Another, potentially complementary influence is the relative width of the HC and this area of dry land at the equator. The HC widens as $T_r$ increases, as is well-known from theoretical and numerical considerations \cite[e.g.][]{Hide1969, Schneider1977, Held_Hou80, Schneider1987, Lindzen1988, Plumb1992, Emanuel1994, Emanuel1995, Caballero2008, Kaspi2015, Guendelman2018, Hill2019, Hill2020}. With constant $\Delta\varphi$, this can allow the HC to ``reach'' the edge and provide a moisture source to the equator. 

This can be used to explain the results shown in Figures \ref{rot_exp_sh_vsh}(g), \ref{rot_exp_sh_vsh}(h), and \ref{rot_exp_rh}. The plots appear to show an unexpected paradox, that being smaller $T_r$ correlates with meeting more Titan-like criteria. As mentioned previously, Titan is a slow rotator, with one Titan-day equaling approximately 16 Earth-days ($T_r=16$). Yet our experiments with variable $T_r$ show that fast rotators (small $T_r$) are more likely to meet our Titan-like criteria. Applying the concept that larger $T_r$ corresponds to a larger HC, we would in fact predict that experiments with smaller $T_r$ and similar $\Delta\varphi$ would be drier and thus more likely to meet our Titan-like criteria. This means that Titan itself must have a competing effect on its equatorial climate. This may be a large $\Delta\varphi$, since if $\Delta\varphi$ is also increased rather than held constant, then it would compete with the wider HC that comes with a larger $T_r$.

\begin{figure}[h]
\includegraphics[width=0.98\textwidth]{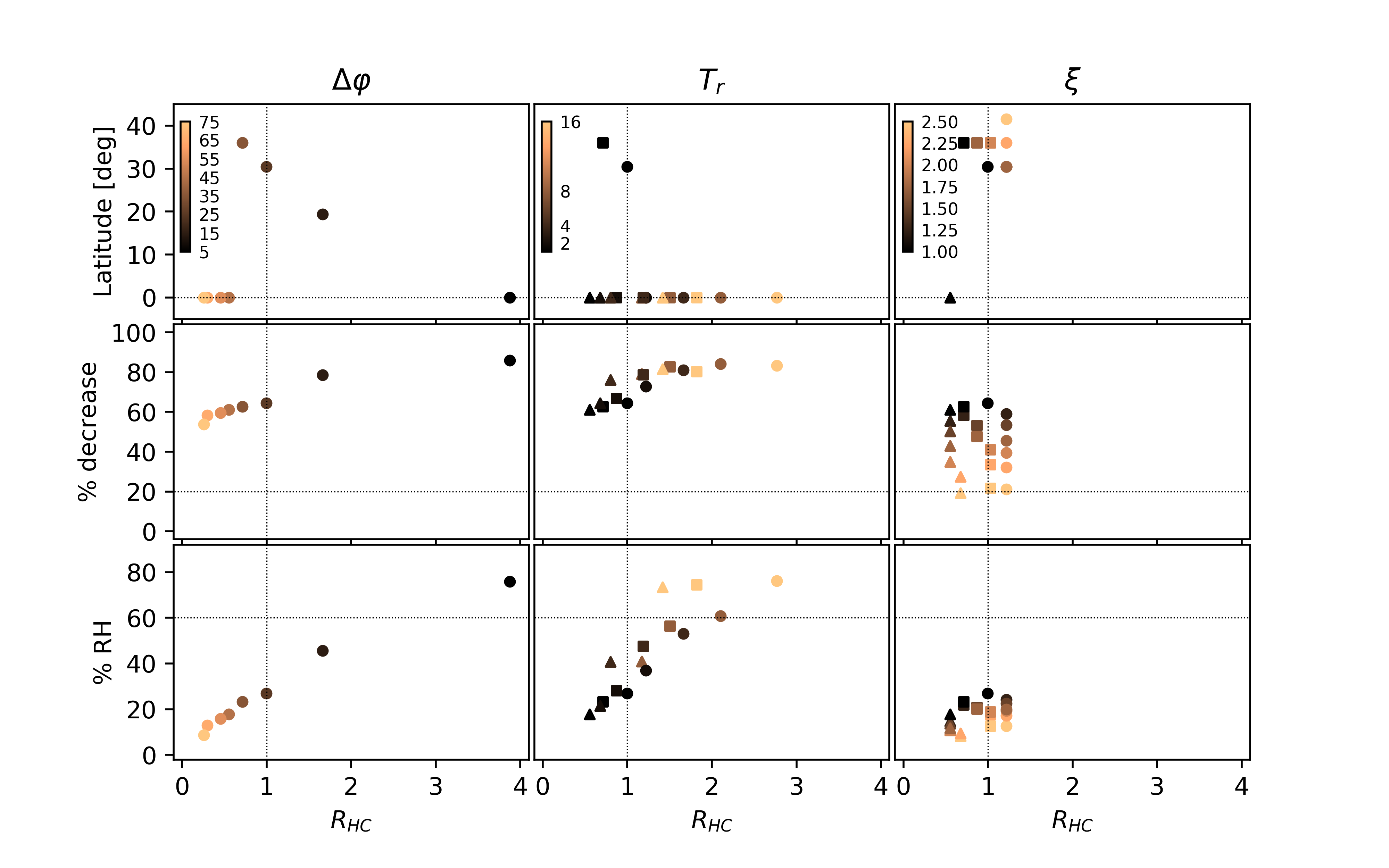}
\caption{Ratio of time-mean HC width to $\Delta\varphi$, $R_{HC}$, vs: (Top row) Latitude of peak zonal- and time-mean $q_1$; (Middle row) percent decrease in time-mean specific humidity at the equator between the surface and 60hPa pressure level; (Bottom row) time-mean RH at the equatorial surface. Left column shows data for the $\Delta\varphi$ experiments; Middle column shows data for the $T_r$ experiments; and Right column shows data for the $\xi$ experiments. The marker colors show the respective paramter values of each experiment set. Dashed horizontal lines denote the relevant criteria thresholds (Top: OffEq, Middle: ConVQ, Bottom: LowRH). Dashed vertical lines denote the 1:1 ratio, below which the HC is narrower than the land strip and thus cutoff from the ocean. Symbols in the middle and right columns have the values of $\Delta\varphi=25^\circ$ (circles), $35^\circ$ (squares) and $45^\circ$ (triangles).}
\label{hcwidth}
\end{figure}

We quantify the ratio between the HC width (defined as the latitude where the zonal-mean mass flux first drops below 10\% of its maximum value while at the level of the maximum value) to $\Delta\varphi$ as $R_{HC}$, shown in Figure \ref{hcwidth}. For $R_{HC}<1$ the HC is contained by the land strip, while for $R_{HC}\geq1$ the HC reaches the shoreline. Each row shows one of the three humidity quantities used to determine the Titan-like criteria as a function of $R_{HC}$. The left column shows data points for the $\Delta\varphi$ experiments, the middle column shows the $T_r$ experiments, and the right column shows the $\xi$ experiments. Each set of markers is colored based on its experimental control parameter value, as indicated in color bars in the top row. Overall, there is a general trend toward meeting more Titan-like criteria with smaller $R_{HC}$, with some caveats. For the $\Delta\varphi$ experiments, the peak latitude of $q_1$ (Figure \ref{hcwidth}, top-left) is off the equator for $R_{HC}$ values between $\sim0.75$ and $\sim2$. Despite having HC widths that are smaller than the land strip width ($R_{HC}$<1), the experiments with the largest $\Delta\varphi$ have their peaks at the equator. This apparent paradox is due to the high latitude of their shorelines and correspondingly lower mid-latitude values of $q_1$ outpacing any reduction in $q_1$ at the equator (see Figure \ref{lat_exp_sh}(a)). For the $\Delta\varphi$ experiments, the percent decrease in the vertical profile of specific humidity (Figure \ref{hcwidth}, middle-left) and equatorial RH (Figure \ref{hcwidth}, bottom-left) both increase with larger $R_{HC}$ and do not have a clear transition at $R_{HC}=1$. Experiments with varying rotation period, $T_r$, have only two outlier cases with off-equatorial peaks in $q_1$ around $R_{HC}=1$ (Figure \ref{hcwidth}, top-middle); for all others, the peak in $q_1$ stays at the equator. For the $T_r$ experiments, the percent decrease in SH (Figure \ref{hcwidth}, middle) and equatorial RH (Figure \ref{hcwidth}, bottom-middle) follow similar patterns to the corresponding panels for the $\Delta\varphi$ experiments. There is a similar range in $R_{HC}$ between the $\Delta\varphi$ and $T_r$ experiments, even though the latter only use three values of $\Delta\varphi$ ($25^{\circ}$, circles; $35^{\circ}$, squares; and $45^{\circ}$, triangles). This shows the widening of the HC with increasing $T_r$ is playing the dominant role in determining $R_{HC}$. In contrast with $\Delta\varphi$ and $T_r$, $R_{HC}$ varies only slightly, from roughly 0.5 -- 1.25, in the $\xi$ experiments. The peak in $q_1$ (Figure \ref{hcwidth}, top-right) is off the equator for all but one experiment, with that experiment having $R_{HC}$ close to 0.5 and the others having larger values. Excluding this outlier, all experiments meet the OffEq and LowRH criteria. The percent decrease in specific humidity (Figure \ref{hcwidth}, middle-right) and equatorial RH (Figure \ref{hcwidth}, bottom-right) show little dependence on $R_{HC}$, the scatter being almost entirely from the range of $\Delta\varphi$ used for these experiments. In contrast, increasing $\xi$ decreases the percent decrease in specific humidity (Figure \ref{hcwidth}, middle-right) without changing $R_{HC}$; only the highest $\xi$ values meet the ConVQ criterion. This suggests that $\xi$ does not significantly affect the extent of the HC, despite affecting the apparent shape of the circulation (Figure \ref{mflux}). 

To summarize, $R_{HC}$ is moderately predictive of meeting our Titan-like criteria for the $\Delta\varphi$ and $T_r$ experiments, but not at all for the $\xi$ experiments. Figure \ref{hcwidth} confirms our previous analysis, that large values of $\xi$ associated with Titan's abundant methane vapor are likely responsible for Titan's nearly uniform SH in the lower troposphere. It also makes clear the competing effects of Titan's mid- or high-latitude shorelines and its wide HC on meeting the OffEQ criterion. For fixed $T_r$, Titan's smaller radius would make its HC fractionally wider than it is on Earth, i.e. it would extend to higher latitudes and exacerbate the problem. Titan's surface liquids would need to be tightly constrained to the very highest latitudes, which given the colder conditions there would make it even more difficult for $q_1$ at the shoreline to exceed even a small bump in equatorial $q_1$. A decent approximation is $R_{HC}\sim 1$ on Titan, and while there are a number of our simulations with $R_{HC}\sim 1$ that meet at least two criteria, only the very largest $\xi$ values meet all three. So it seems that while the relative widths of Titan's dry equatorial belt and HC width are important, it is the abundant methane that is key to establishing our three Titan-like criteria.

\section{Conclusions}

We have presented a suite of idealized GCM experiments of an Earth-like climate while varying four parameters to investigate the transition to a Titan-like climate at the equator. We created three criteria to determine if the equatorial climate was Titan-like: (1) The peak in $q_1$ is not at the equator, despite the peak in near-surface temperature being there; (2) The specific humidity is approximately constant through the lower troposphere at the equator; (3) The annual-mean near-surface RH at the equator is below 60\%, indicating a climate that is significantly moisture limited. The first parameter we varied is the width of a land strip centered on the equator that was initialized using a novel model of land hydrology \citep{faulk2020titan}. We found that varying the land strip ($\Delta\varphi$) alone did not allow any experiments to meet all three criteria. We then varied the rotation ($T_r$) for experiments with $\Delta\varphi$ between 25-$45^{\circ}$. Similarly to the experiments with only a varied land strip, we found that varying $T_r$ did not meet all three criteria. Notably, the trend is such that lower $T_r$ created more Titan-like conditions, which is in contradiction to Titan's high $T_r$ relative to Earth. This apparent contradiction will require further investigation, most likely by varying $T_r$ and $\xi$ in tandem. The fourth control parameter of our experiments is the volatility of the condensable ($\xi$), which we varied for the same land strip widths as with the $T_r$ experiments. We found that varying $\xi$ did achieve all three criteria for $\xi=2.5$. Titan's condensable, methane, is much more volatile than water on Earth, so this finding indicates that the volatility itself may have a significant influence on Titan's low-$q_1$ climate. We also found a competition between $\Delta\varphi$ and $\xi$, in that wider land strips were less Titan-like for the same $\xi$. Our $\xi$ experiments only go up to $\Delta\varphi=45^{\circ}$, which is likely narrower than the desert region on Titan. To compensate, the effective $\xi$ on Titan may be higher than the maximum $\xi$ used in our experiments, potentially around 3. While it is feasible to study larger $\Delta\varphi$, we are limited to a maximum $\xi$ because of the water vapor feedback, and expanding the range in $\xi$ will require an adjustment to the radiative transfer. We will explore this in future work.

Our experiments identify several outstanding questions.
\begin{enumerate}
\item Can larger $T_r$ allow us to achieve higher values of $\xi$ in our model for the same land-strip widths?
\item How high does $\xi$ need to be to achieve Titan-like conditions on a land strip that covers a similar area to Titan's large equatorial desert?
\item Will varying three parameters ($\Delta\varphi$, $T_r$, and $\xi$) simultaneously be necessary to create a more Titan-like climate?
\end{enumerate}

In addition to understanding a Titan-like climate, these experiments can be used to shed light on Earth-like climates, including our own planet's past. Earth has had several continental arrangements through its history, and our hydrology module can be used to investigate them in detail by specifying global topography. As demonstrated by our experiments with a land strip of varying width, the presence, location, and size of a continent can have significant effects on global climate. With the addition of a seasonal cycle and more complex continental arrangements, many points in Earth's history can be simulated and compared to provide better understanding of the range of possible climate states.

\acknowledgments
MMM and JLM acknowledge support from NSF award 1912673. 

\datastatement
Data for all experiments will be made available on Zenodo.

\clearpage

\bibliographystyle{ametsocV6}
\bibliography{refs}

\end{document}